\pdfoutput=1
\documentclass[preprint,showpacs,preprintnumbers,amsmath,amssymb]{revtex4}

\usepackage{graphicx}% Include figure files
\usepackage{dcolumn}% Align table columns on decimal point
\usepackage{bm}% bold math

\begin{document}

\preprint{APS/123-QED}

\title{First principles calculations of the electronic and geometric structure of $Ag_{27}Cu_{7}$ Nanoalloy}% Force line breaks with \\

\author{Marisol Alc{\'a}ntara Ortigoza}
\email{alcantar@physics.ucf.edu}
\affiliation{Department of Physics, University of Central Florida\\
Orlando, Florida 32816-2385, USA
}%

\author{Talat S. Rahman}
\email{talat@physics.ucf.edu}
\affiliation{Department of Physics, University of Central Florida\\
Orlando, Florida 32816-2385, USA
}%

\date{\today}% It is always \today, today,
             %  but any date may be explicitly specified

\begin{abstract}
\emph{Ab initio} calculations of the structure and electronic density of states (DOS) of the perfect core-shell $Ag_{27}Cu_{7}$ nanoalloy attest to its $D_{5h}$ symmetry and confirm that it has only 6 non-equivalent (2 $Cu$ and 4 $Ag$) atoms. Analysis of bond-length, average formation energy, heat of formation of $Ag_{27}Cu_{7}$ and $L1_2$ $Ag-Cu$ alloys provide an explanation for the relative stability of the former with respect to the other nanoalloys in the same family. The HOMO-LUMO gap is found to be 0.77 eV, in agreement with previous results. Analysis of the DOS of $Ag_{27}Cu_{7}$, $L1_2$ $Ag-Cu$ alloys and related systems provides insight into the effects of low coordination, contraction/expansion and the presence of foreign atoms on the DOS of $Cu$ and $Ag$.
While some characteristics of the DOS are reminiscent of those of the phonon-stable $L1_2$ $Ag-Cu$ alloys, the $Cu$ and $Ag$ states 
hybridize significantly in $Ag_{27}Cu_{7}$, compensating the $d$-band narrowing that each atom undergoes and hindering the dip in the DOS found in the bulk alloys. Charge density plots of $Ag_{27}Cu_{7}$ provide further insights into the relative strengths of the various interatomic bonds. Our results for the electronic and geometric structure of this nanoalloy can be explained in terms of length and strength hierarchies of the bonds, which may have implications also for the stability of alloy in any phase or size.
\end{abstract}

% PACS, the Physics and Astronomy
                             % Classification Scheme. Use showpacs class option if pacs is desired
%\keywords{Bimetallic nanoalloys, Nanoclusters, Alloys lattice dynamics, Electronic structure, charge density}

%61.46.Bc Structure of clusters (e.g., metcars; not fragments of crystals; free or loosely aggregated or loosely attached to a substrate) (see also 61.48..c for structure of fullerenes) 
%61.46.Df Structure of nanocrystals and nanoparticles ("colloidal" quantum dots but not gate-isolated embedded quantum dots)  
%61.66.Dk Alloys  
%63.20.dk Phonons in crystal lattices  First-principles theory 
%71. Electronic structure of bulk materials  
%71.15.Nc Total energy and cohesive energy calculations  
%71.55.Ak Metals, semimetals, and alloys  
%71.55.Ak Metals, semimetals, and alloys  
\maketitle

\section{\label{sec:level1}introduction}

Small bimetallic nanoclusters often have physical and chemical properties that are distinct from that of their pure bulk counterparts and suggestive of novel applications.~\cite{a1,a2,a3} Not surprisingly, materials assembled from finite-sized bimetallic clusters have been investigated intensively not only for their catalytic and optical properties,~\cite{a4,a5,a6,a7,a8,a9,a10,a11,a12,a13,a14} but also for their ability to assemble into \emph{cluster crystals}~\cite{a15,a16} and their possible applications in single-electron tunnelling devices.~\cite{a1}
Along with its high symmetry and relatively high melting temperature, one of the criteria for a cluster to be used as a potential building block for cluster-assembled materials is its chemical stability relative to other reagents and to other clusters of the same material. Also, major difficulties arise from the fact that clusters may tend to coalesce when assembled. This can be prevented in one of two ways - either by isolating the clusters in matrices or by coating them with surfactants.~\cite{a16}
An alternative route is to find nanoclusters that are naturally stable, i.e., nanoclusters whose intra-cluster interaction is stronger than the inter-cluster interaction allowing the clusters to keep their individual identity intact upon assembling. Even so, cluster-assembled materials could still be metastable against dissociation into their bulk phases.

Darby et al.~\cite{a3}, using many-body Gupta potentials, studied the structure and stability (as reflected by the total energy) of a wide variety of $Cu_{x}Au_{y}$ nanoclusters with up to 56 atoms and $x/y = 1, 3$; corresponding to the well-known $x:y$ ratios that result in stable ordered bulk phases at low-temperatures.~\cite{a17}  They found that the geometry of the cluster is influenced by the tendency to maximize the number of $Cu-Au$ and $Au-Au$ bonds. Rossi et al.~\cite{a1}, on the other hand, proposed a new family of 34-atom bimetallic alloys using the genetic global optimization technique (GGO).~\cite{a1}
These nanoalloys are characterized by a perfect core-shell structure in which the \emph{smaller} atoms ($Cu$ or $Ni$) compose the core whereas, the relatively larger, $Ag$ atoms lie on the surface. They find the nanoalloys to be energetically and thermodynamically more stable than pure clusters $Ni_{34}$, $Cu_{34}$, and $Ag_{34}$,~\cite{a1}  and attribute the relative stability of the nanoalloy structures to the supplanting of the inner $Ag$ atoms by \emph{smaller} atoms ($Cu$ or $Ni$) thereby reducing the internal strain in $Ag_{34}$, or the replacement of outer $Cu$ atoms by \emph{larger} atoms ($Ag$) to reduce the external strain in $Cu_{34}$.
As we shall see later, the most stable of the 34-atom $Ag-Cu$ nanoalloy family proposed by Rossi et al,~\cite{a1} $Ag_{27}Cu_{7}$, provides a hint that it is not the $x:y$ ratio that guarantees the stability of either bulk or nano-alloys, rather it is the maximization of the number of \emph{optimized} $Cu-Cu$ and $Cu-Ag$ bonds.

Among a set of possible core-shell nanoclusters modelled by many-body interatomic potentials, Rossi et al.~\cite{a1}  chose the compositions corresponding to their most stable structures for some selected sizes and \emph{locally} optimized the structures using density functional theory (DFT) to confirm the trends given by the GGO and to single out the clusters with high electronic stability, namely, with the largest width of HOMO-LUMO (Highest Occupied Molecular Orbital - Lowest Unoccupied Molecular Orbital) gap. They
determined the thermodynamic stability of the chosen structures through calculations of the melting temperatures from molecular dynamics simulations and of temperature-dependent probabilities of the global minima by harmonic thermodynamics.~\cite{a18} Among the 34-atom family of nanoclusters, they found $Ag_{27}Cu_{7}$ and $Ag_{27}Ni_{7}$  to have the least excess energy with respect to bulk atoms (the lowest heat of formation), strong electronic stability (large HOMO-LUMO gap), and relatively high melting temperatures.~\cite{a1} For these compositions, the GGO method finds the lowest energy structure to have $D_{5h}$ symmetry in which the 7 core-$Cu$ atoms form a decahedron  while the 27 shell-$Ag$ atoms are placed in an anti-Mackay overlayer. In general, the formation of nanoclusters is driven by the tendency to minimize dangling bonds.~\cite{a1} Even though such a tendency would lead to spherical structures,~\cite{a1} the presence of $d$-states increase the tendency to create directional bonding ($s$-$d$ hybridization), which is manifested by the preference for well-ordered atomic structures, such as the decahedral forms.~\cite{Issendorf}

Experiments and heat of formation calculations have shown that $Ag-Cu$ alloys generally tend to segregate.~\cite{a34,a35,a36} In a sense, the core-shell structure of $Ag_{27}Cu_{7}$ nanoalloy is itself segregated. In order to understand how its particular geometry implicitly stabilizes it, a detailed examination of the relative strengths and lengths of the $Ag-Cu$, $Cu-Cu$, and $Ag-Ag$ bonds is needed. For bulk $Cu-Au$ alloys the presence of a dip in the electronic DOS near the Fermi level was also considered to be a signature of alloy stability.~\cite{a34}  Interestingly, such a dip is not found in bulk $Au-Ag$ alloys.~\cite{a17,a34} Moreover, in the case of crystalline solids,
the structural stability is linked to the absence of phonon instabilities.~\cite{r1} The purpose of this paper is to carry out a full investigation of the relationship between the geometric and electronic structure of $Ag_{27}Cu_{7}$ nanoalloy and related bulk systems to get insights into the various factors that may impact their stability. That is,
through examination of the formation energy, the density of states near the Fermi level, the HOMO-LUMO gap, and the charge density distribution, we develop criteria which may lead to chemical and electronic stability of $Ag_{27}Cu_{7}$. To obtain additional insights into the structure-stability relationship, we have also carried out calculations of the bond-length, the electronic structure, and the phonon dispersion of $Ag_{3}Cu$  and $Cu_{3}Ag$  bulk alloys in their $L1_{2}$  phase.~\cite{naval}

The rest of the paper is organized as follows: Section~\ref{acII} contains the computational details, while Section~\ref{acIII} is a summary of our results and is divided in subsections \ref{acIIIA} and \ref{acIIIB}. Subsection \ref{acIIIA} analyzes the geometry and bond coordination of the atoms in the $Ag_{27}Cu_{7}$ nanoalloy and those in the bulk $Ag-Cu$ $Ll_{2}$ alloys, for insights into proposed stability criteria. In Subsection \ref{acIIIB}, we examine our calculated DOS of $Ag_{27}Cu_{7}$ and of the bulk systems ($Cu_{3}Ag$ and $Ag_{3}Cu$ ), and the charge density distribution of $Ag_{27}Cu_{7}$. Finally, in Section \ref{acIV} we summarize our conclusions, and discuss how the relation between structure and stability in $Ag_{27}Cu_{7}$ nanoalloy can be understood in terms of a specific hierarchy in bond strength and the capability of the structure to provide the bond lengths for which that hierarchy is satisfied.

\section{Computational Details}\label{acII}

Periodic super-cell calculations are performed in the framework of density functional theory.~\cite{r50} Our calculations are based on the pseudopotential approach and the plane wave method (Quantum ESPRESSO: opEn-Source Package for Research in Electronic Structure, Simulation, and Optimization).~\cite{r57}
Ultra-soft pseudo-potentials~\cite{r53} used here are generated consistently with GGA schemes. For the GGA functional the expression introduced by Perdew, Burke, and Ernzerhof (PBE) has been introduced.~\cite{r56} Integrations up to the Fermi surface are performed by using a broadening technique~\cite{a25}  with smearing parameter of 0.2 eV (0.147 Ry). Below we provide some specifics of the calculations as relevant to a particular system.

\subsection{Calculation of bulk systems}\label{acIIA}

We have performed extensive convergence tests for lattice constants, bulk moduli and total energies of bulk $Cu$, $Ag$, $Ag_{3}Cu$ , and $Cu_{3}Ag$ . To obtain the minimum energy configuration with zero stress based on total energy differences of 1 mRy, as well as, convergence up to the third and second digit in the lattice parameter (in a.u.), and the bulk modulus (in Mbar), respectively, the calculations demand a k-point sampling of 145 Monkhorst-Pack special points~\cite{a26} (corresponding to a $16 \times 16 \times 16$ Monkhorst-Pack grid) for the integrations over the Brillouin zone (BZ). Furthermore, the plane wave kinetic energy cut-off, $E_{cut}$, and the energy at
which the charge density Fourier expansion is truncated, $E_{\rho}$, had to be set equal to 680 eV (50Ry) and 8160 eV (600 Ry), respectively. These convergence criteria surpass by far most of those reported in the literature, but are necessary to obtain reliable results in the present case, as we will see.

The lattice dynamics of $Ag_{3}Cu$  and $Cu_{3}Ag$  bulk alloy at arbitrary wave-vectors is obtained by the Perturbational DFT (DFPT) which is based on the linear response theory.~\cite{r62,r63,r65} To determine the force constants we use a $2 \times 2 \times 2$ q-point mesh in the BZ of the $Ll_{2}$  structure.~\cite{r63} Phonon dispersion curves are obtained by the standard Fourier interpolation method.~\cite{r63}

\subsection{Calculation of $Ag_{27}Cu_{7}$ nanoalloy and isolated atoms}\label{acIIB}

Since in the unrelaxed configuration of $Ag_{27}Cu_{7}$ the separation between the most distant atoms is about 8.7 \AA, we locate the nanoalloy inside a cubic super-cell with side length of 24 \AA. In this manner we ensure that as a result of periodic boundary conditions, the atoms at the edges of neighboring clusters are at least 15 {\AA}  apart, thereby isolating the clusters from each other. The same cubic box is used to model and calculate the total energy of isolated $Cu$ and $Ag$ atoms, using a spin-polarized calculation.

In this work the Broyden-Fletcher-Goldfarb-Shanno (BFGS) algorithm~\cite{r70} is used to minimize the nanoalloy total energy as a function of atomic positions.~\cite{r57} At equilibrium, forces on the nanocluster atoms are required to be below 6.5$\times$10$^{-4}$ eV/{\AA} (2.6$\times$10$^{-5}$ Ry/au). 
Given the large separation between nanoclusters (isolated atoms) in neighboring supercells, integrations over the BZ using only 1 k-point lead to reliable convergence in the calculated values of the total energy (see Ref.~\onlinecite{comment}).
For the nanoalloy, $E_{cut}$ and $E_{\rho}$ are 680 eV and 8160 eV, respectively, as mentioned above. Since these parameters surpass the demands for convergence in other systems involving copper or silver, and using the same DFT code,~\cite{r80,a32} we expect them to work well for the nanoalloys of interest here. While these demands for convergence make the calculations very cpu intensive, it is worth mentioning that we find the total energy and even the geometry of the nanoalloy to be severely affected if we were to use the default values of $E_{cut}$ = 340 eV and $E_{\rho}$ = 1360 eV in the code. For example, with the latter choice, the fully relaxed $D_{5h}$ structure (resulting from $E_{cut}$ = 680 eV and $E_{\rho}$ = 8160 eV) is no longer stable and relaxes towards a structure in which the atomic positions break the $D_{5h}$ symmetry by displacements of up to 1 {\AA} from their original positions and forces cannot be lowered below $2.5 \times 10^{-2}$ eV/{\AA}.

\section{Results and Discussion}\label{acIII}

Since the nanoalloy presents several opportunities for comparison of its properties and development of criteria for its stability, we find it beneficial to divide this section into two subsections, each of which consists of several parts. In subsection~\ref{acIIIA}, we concentrate on issues related to the geometry, the distribution of bond lengths, the atomic coordination, and the formation energy of the nanoalloy. We first introduce a notation in subsection~\ref{acIIIA1} that classifies the atoms in $Ag_{27}Cu_{7}$ according to their location within the nanoalloy. In~\ref{acIIIA2}, we inspect how the local coordination of the atoms in the nanoalloy
relates to the bond lengths. Since there is hardly any experimental data on $Ag_{27}Cu_{7}$ and since calculations of the phonon density of states of the nanoalloy from first principles is still a challenge, we have included in subsection~~\ref{acIIIA3} our results for the structure and dynamics of bulk alloys, $Ag_{3}Cu$  and $Cu_{3}Ag$, to gain insights and draw stability criteria for the nanoalloy of interest here. The average formation energy of $Ag_{27}Cu_{7}$ nanoalloy and stability considerations are analyzed in subsection~~\ref{acIIIA4}.

In subsection~\ref{acIIIB} we focus on the electronic DOS and the local charge density of the nanoalloy. Subsection~\ref{acIIIB1} contains the electronic DOS of the $Ag_{27}Cu_{7}$ nanoalloy and includes for comparison also those of bulk alloys, $Ag_{3}Cu$  and $Cu_{3}Ag$. The local charge density distribution in the nanoalloy is summarized in subsection~\ref{acIIIB2}.

\subsection{Geometry, bond coordination, and stability considerations}\label{acIIIA}

\subsubsection{Geometric Structure of $Ag_{27}Cu_{7}$ nanoalloy}\label{acIIIA1}

The initial configuration we adopted for $Ag_{27}Cu_{7}$~\cite{a19}  nanoalloy nicely relaxes towards the $D_{5h}$ symmetry after energy minimization, as seen from the plots in Fig.~\ref{fig:ac1}. Accordingly, there are only 6 types of non-equivalent atoms: 4 types of $Ag$ and 2 types of $Cu$ atoms. This leads to a natural and useful classification of the atoms that refers to their distance from the mirror plane: $Cu$ layer 0 ($Cu0$), $Ag$
layer $0$ ($Ag0$), $Cu$ layer $\pm 1$ ($Cu1$), $Ag$ layer $\pm 2$ ($Ag2$), $Ag$ layer $\pm 3$ ($Ag3$), $Ag$ layer $\pm 4$ ($Ag4$), as shown in Fig.~\ref{fig:ac1}. Layer $0$, which lies on the mirror plane, consists of two pentagonal structures; the smaller one is made of copper atoms ($Cu0$) and fits in the larger one that is composed of silver atoms ($Ag0$) (Fig.~\ref{fig:ac1}(a) and (b)). The \emph{single-atom layers}, layers $\pm 1$ and $\pm 4$, sit on the 5-fold rotation axis (see Fig.~\ref{fig:ac1}(c) and (f)). The other two pentagonal layers, $\pm 2$ and $\pm 3$, are centered at the 5-fold symmetry axis (Fig.~\ref{fig:ac1}(d) and (e)). The radii of $Ag2$ and $Cu0$ pentagons are parallel to each other, but rotated $36^{\circ}$ with respect to the $Ag0$ and $Ag3$ pentagons, as shown in Fig.~\ref{fig:ac2}.
Since the layers are symmetric with respect to a mirror plane, the cluster can be characterized by only five of them, say, the central layer (layer $0$) and those above this (layers $1$ to $4$). Ultimately, the symmetry of $Ag_{27}Cu_{7}$ allows us to fully describe its geometric structure by 8 parameters: the interlayer distances ($d_{01}$ = 1.341, $d_{12}$ = 0.052, $d_{23}$ = 0.992, $d_{34}$ = 1.512 {\AA}) and the pentagons. side length ($a_{Cu0}$ = 2.584, $a_{Ag0}$ = 4.897, $a_{Ag2}$ = 5.115, $a_{Ag3}$= 2.948 {\AA}, as shown in Fig.~\ref{fig:ac2}). The dislocations in the $D_{5h}$ structure, as present in our initial configuration,~\cite{a19} relaxed into the perfect $D_{5h}$ structure under the stringent criterion that the apothem of each pentagon can be well defined up to 0.0003 {\AA}, while interlayer distances are well defined up to 0.0001 {\AA}.
The interatomic bond lengths for the 34 atoms in the cluster, in the relaxed geometry, are summarized in Table~\ref{tab:table1}, and discussed in detail below.

\subsubsection{Neighbor distances in $Ag_{27}Cu_{7}$: a comparison with bulk $Ag$ and bulk $Cu$ values}\label{acIIIA2}

Despite the perfect $D_{5h}$ symmetry of the cluster, Table~\ref{tab:table1} shows an intricate hierarchy of bond lengths in the optimized structure. For example, $Ag3$ atoms have 9 neighbors spread in a range from 2.726 to  2.948 {\AA}, all within a separation equal to the bond length of $Ag$ bulk. Figs.~\ref{fig:ac3} (a) and (b) show a comparison between the bond coordination of atoms in the $Ag_{27}Cu_{7}$ nanoalloy and that of atoms in bulk $Ag$ and bulk $Cu$. Notice from Table~\ref{tab:table1} that the local coordination of $Cu$ atoms in $Ag_{27}Cu_{7}$ seems, in fact, not dramatically different from that in bulk $Cu$ regarding the number of first and second
NN. The $Cu0$ atoms, for example, have 12 neighbors within a distance of 2.548 and 2.786 {\AA}, and another twelve between 4.181 and 4.482 {\AA}.  Silver atoms, on the other hand, find themselves in unusual atomic environments: at distances ~3 {\AA} (bulk $Ag$ nearest neighbors distance) $Ag2$ and $Ag4$ atoms have barely acquired 6 neighbors, while $Ag3$ and $Ag0$ atoms get 9 and 8 NN, respectively. Between 4.2-4.5 {\AA}, at which bulk $Ag$ atoms already have 18 NN, $Ag0$, $Ag3$ and $Ag4$ atoms have acquired at most 11 neighbors, while $Ag2$ atoms have only 7 NN, finding themselves as the most under-coordinated atoms of the cluster.
Most importantly, we notice that the NN bond lengths between $Cu$ atoms contract by as much as 2{\%} with respect to the value in bulk (2.599 {\AA}). Such contraction may be expected for $Cu$ atoms in low coordination environments, such as on the surface
layer of $Cu(100)$ 
(inward contraction of around 3~{\%}~\cite{r1}), and not in $Ag_{27}Cu_{7}$ because of their high coordination (12)
as in bulk $Cu$. Interestingly, for the shell $Ag$ atoms, which have much lower local coordination than the core $Cu$ atoms, the $Ag-Ag$ NN bond lengths are at most $\sim$ 2.6 {\%} shorter than those in bulk $Ag$ (2.943 {\AA}). If coordination alone were a measure, one might have expected the most under-coordinated $Ag$ atoms, $Ag2$ and $Ag4$, to undergo a larger contraction, as found on $Ag$ surfaces.~\cite{r1} Instead, in $Ag_{27}Cu_{7}$, $Ag2$ and $Ag4$ form short bonds of about 2.59 {\AA} with their neighboring $Cu$ atoms, as seen from Table~\ref{tab:table1}. These bond lengths are in fact very close to the smallest $Cu-Cu$ bond lengths (2.55 {\AA}) and considerably smaller than the smallest $Ag-Ag$ bond lengths (2.87 {\AA}). Indeed, it follows from Table~\ref{tab:table1} that in $Ag_{27}Cu_{7}$ the first NN of every $Ag$ atom is a $Cu$ atom, pointing to the reality that finite sized structures of these elements may not follow straightforwardly the behavior of infinite and/or semi-infinite systems and the
relationship between bond coordination number and bond stiffening might be subtler in nanoalloys. Two conclusions may, nevertheless, be drawn from the above: 1) the fact that $Cu-Cu$ bond lengths in $Ag_{27}Cu_{7}$ contract almost as much as $Ag-Ag$ bond lengths suggests that $Cu$ atoms are more sensitive than $Ag$ atoms to either low coordination, or local geometry, or chemical environment; 2) the low coordination of the $Ag$ atoms appears to be significantly compensated by the formation of short bond lengths with $Cu$ atoms.
To discriminate between the above mentioned effects of local coordination, geometry, and environment,
the conclusions about bond lengths in the $Ag_{27}Cu_{7}$ nanoalloy need to be put on firmer grounds through examinations of details of the electronic structure and the charge density distribution, and their implications for nanoalloy stability. We will turn to this in Sections~\ref{acIIIA4}  and \ref{acIIIB}. But before we do that, it is interesting to examine the
structural stability of related bulk alloys, $Ag_{3}Cu$  and $Cu_{3}Ag$, for 
which some information already exists and hence can serve as reference points. The relationship between the composition of $Ag_{27}Cu_{7}$ and $Ag_{3}Cu$  is obvious. The other alloy is chosen to establish whether preponderance of $Cu$ and/or its effect on bond
lengths is a key for understanding the structural stability of these alloys.

\subsubsection{Structure, phonons, and heat of formation of $Cu_{3}Ag$  and $Ag_{3}Cu$  bulk alloys}\label{acIIIA3}

In considerations of structural stability of bulk alloys, it is essential that the heat of formation be negative and that the phonon spectrum be well defined. $Ag-Cu$ alloys, unlike $Au-Cu$ and $Au-Ag$ alloys, are known for their tendency to segregate and have a miscibility gap beyond the eutectic temperature of the material.~\cite{a34,a35,a36}  Earlier calculations~\cite{a34,a35,a36} have shown that $Ag-Cu$ alloys possess a positive heat of formation regardless of the chosen ratio of $Ag$ to Cu.~\cite{a34,a35}  In particular, the heat of formation per atom was found to be about 70 meV for $Ag_{3}Cu$  (60 meV, in this work) and 80 meV for $Cu_{3}Ag$  (64 meV, in this work), pointing to the structural instability of these $Ag-Cu$ bulk alloys.~\cite{a35}

It has, however, been pointed out that even with a positive heat of formation presence of a well-defined phonon spectrum may serve as an indicator of alloy stability under special formation conditions, for example, using non-equilibrium techniques.~\cite{a38,a39,a40,a41}  Indeed, among a variety of immiscible noble-transition metal alloys, some - particularly those that mix fcc and hcp metals - have shown mutual solid solubility.~\cite{a42} Kong et al.~\cite{a38} have calculated the phonon spectra of several structures of the equilibrium immiscible $Ag_{x}Ru_{y}$ alloys, for $x/y$ = 1/3 and 3, to find that only the $L1_{2}$ and $D0_{19}$ phases of $Ru_{3}Ag$ may be stable. In Ref.~\cite{a38}, phonon-stable/unstable structures were associated with the presence of relatively high charge density bridging NN atoms of the same/different element. We will come to charge density implications later in Section~\ref{acIIIB2}. For the present discussion, we note that, regardless of the structure of the stable phases, the
phonon-stable $Ag-Ru$ alloys are all $Ru$-rich and have smaller lattice parameter than the corresponding $Ag$-rich structures, owing to the fact that typical bond lengths in bulk $Ru$ are smaller than those in bulk $Ag$. Moreover, Kong et al. were able to remove the phonon instabilities, when present, by artificially increasing the external pressure; i.e., by simply reducing the equilibrium lattice parameter.

In the above spirit, we now turn to the calculation of the structure and the lattice dynamics of $Ag_{3}Cu$  and $Cu_{3}Ag$  bulk alloys using DFT and DFPT methods. We find the bond-length for $Ag_{3}Cu$  to be 2.87 {\AA} and that of $Cu_{3}Ag$  to be 2.70 {\AA}. For reference, note that our calculated bond lengths for bulk $Cu$ and
$Ag$ are 2.59 and 2.94 {\AA}, respectively. Note also that the shortest $Ag-Cu$ bond lengths in $Ag_{27}Cu_{7}$ are around 2.6 {\AA}. The bond-lengths in bulk alloys $Ag_{3}Cu$  and $Cu_{3}Ag$  are thus larger than the shortest $Ag-Cu$ bond-length in the nanoalloy and that in bulk $Cu$ and may  imply lack of overlap of the $d$-orbitals for its $Ag-Cu$ and $Cu-Cu$ bonds (as we shall see), pointing to the structural instability of these bulk alloys. Our calculated phonon dispersion curves (Fig.~\ref{fig:ac4}) of $Ag_{3}Cu$  and $Cu_{3}Ag$, showing the absence of unstable modes, however, suggest that these alloys may be stable and obtainable by non-equilibrium techniques.~\cite{a38}

Phonon dispersion curves are furthermore a measure of the contribution of the vibrational entropy to the free energy of a given system.~\cite{a43} The vibrational entropy integrates the vibrational DOS weighted by a factor that falls off as the frequency of phonons increases. Thus, vibrational entropy plays a larger role in the minimization of the free energy for systems whose (well-defined) phonon dispersion curves display notable
contributions and shifts of the density of states towards the lower frequency range.~\cite{a36} For example, bulk $Cu_{3}Au$  ($L1_{2}$) has larger vibrational entropy than either bulk $Au$ or $Cu$.~\cite{a45} Our calculated phonon dispersion for bulk $Ag_{3}Cu$  (see Figs.~\ref{fig:ac4}(a)) is softer than that of bulk alloy $Cu_3Ag$ (Figs.~\ref{fig:ac4}(b)), bulk $Ag$,~\cite{a44} and bulk $Cu$,~\cite{a44} indicating that the vibrational entropy of $Ag_{3}Cu$  is larger than that of the bulk $Ag$ and $Cu$. Regardless, the vibrational entropic contribution to the reduction of the free energy is small~\cite{a36} (typically less than 5 meV at room temperature) compared to the positive heat of formation ($\sim$60 meV) found for this bulk alloy. Instead, the presence
of stiffer vibrational DOS of $Cu_{3}Ag$, compared to that of $Ag_{3}Cu$, reflects stronger bonds in the former which may provide stability if created by the techniques mentioned above.~\cite{a38,a39,a40,a41}

%For bulk alloys, the configurational entropy may  play a more important role than the vibrational entropy with contributions to the free energy of up to $\sim$~0.7$k_B$.~\cite{a36} In the case of nanoalloys, however, although the number of atomic permutations that lead to the same arrangement can be easily calculated, the number of all possible arrangements of all possible isomers of 34 atoms, even for pure clusters,~\cite{Ag23} is by far much larger in such a way that the configurational energy is expected to have little impact on the free energy of a given nanoalloy. Nonetheless, these issues related to entropic contributions are multifaceted and subtle and need to be the subject of further investigation.

\subsubsection{Formation energy of $Ag_{27}Cu_{7}$}\label{acIIIA4}

In Ref.~\cite{a1}, the thermodynamic stability of a given nanoalloy is evaluated via considerations of its melting temperature and its relative energetic stability is established through comparison of the heats of formation, $\Delta$, adapted to binary clusters as follows:
\begin{equation}\label{eqac2}
\Delta (Ag_{N_1}Cu_{N_2}) = \frac{E(Ag_{N_1}Cu_{N_2}) - N_1 E(Ag_{bulk}) - N_2 E(Cu_{bulk})}{N^{\frac{2}{3}}},
\end{equation}
where $N = N_1+ N_2$,  $E(Ag_{N_1}Cu_{N_2})$ is the total energy of the nanocluster, and $E(Cu_{bulk})$ and $E(Ag_{bulk})$ are the total  energies of one $Cu$ and one $Ag$ atom, respectively, in the bulk phase.
Remarkably, although clusters with increasing binding energies per atom do not necessarily have higher melting temperatures,~\cite{a46} $Ag_{27}Cu_{7}$ came out with both the highest melting point and the least heat of formation. In terms of stability and minimum-energy structures, however, it is important to know also the output given by the average formation energy, which measures the dissociation or \emph{cohesive} energy of the nanoalloy, and to analyze the meaning, implications, and scope of these two energetic considerations. Thus, to estimate the average strength of the bonds, we calculate the so-called average \emph{formation energy per atom}, $E_{form}$, which is defined as,
\begin{equation}\label{eqac1}
E_{form} (Ag_{N_1}Cu_{N_2}) = \frac{E(Ag_{N_1}Cu_{N_2}) - N_1 E(Ag_{free}) - N_2 E(Cu_{free})}{N},
\end{equation}
where $E(Cu_{free})$ and $E(Ag_{free})$ are the energies of isolated $Cu$ and $Ag$ atoms, respectively. We find $E_{form} (Ag_{27}Cu_{7})$ to be 2.17 eV. Since no experimental data on this particular binary nanocluster exists, we turn to the formation energy of related systems for comparison. For example, our calculated
cohesive energy of bulk $Ag$, $Ag_{3}Cu$, $Cu_{3}Ag$, and $Cu$ are 2.51, 2.66, 3.06, and 3.34 eV, respectively, implying that $E_{form}(Ag_{27}Cu_{7})$ is smaller than all these bulk values. Note that the higher cohesive energy of bulk $Ag_{3}Cu$  as compared to that of bulk $Ag$ signals a stronger $Ag-Cu$ bond than the usual $Ag-Ag$ one. In fact, from the results presented in Ref.~\cite{a1}, one finds that the average formation energy per atom of the 34-atom family decreases monotonically from $\sim$2.6 to $\sim$2.0 eV as the $Ag$ content increases from 0 to 34. Considerations of formation energy alone would thus imply that in this family of nanoalloys the $Cu-Cu$ and $Cu-Ag$ bonds are stronger than the $Ag-Ag$ bonds and that $Ag_{27}Cu_{7}$ is not the most stable 
structure. 
It is thus surprising that a related quantity, i.e. the heat of formation (defined as in Eq.~\ref{eqac1}, but substituting $E(Cu_{free})$ and $E(Ag_{free})$ by the cohesive energy of $Cu$ and Ag, respectively), plotted as function of the Ag/Cu ratio in Ref.,~\cite{a1} shows a minimum at this intermediate composition ($Ag_{27}Cu_{7}$) - a result that stands in contrast to that found in $Ag-Cu$ bulk alloys.~\cite{a34,a35,a36} 
Nevertheless, we find that the heat of formation of the $Ag_{27}Cu_{7}$ is 9 times larger than that of $Ag_{3}Cu$, but this suggestion of instability might be misleading since the heat of formation  not only measures the strength of the bonds but also weighs the
energetic cost(gain) of breaking(forming) single element bulk bonds to form(from breaking) binary bonds.
In reality, the formation of intricately tailored structures as $Ag_{27}Cu_{7}$ is not expected to occur simply by melting the parent compounds.
Thus measures like heat of formation have to be supplemented by others such as the dynamical stability of the alloy as displayed by its vibrational modes.
The heat of formation is perhaps more of an indicator of the
life-time of the nanoalloy, say, against clustering and the eventual formation of segregated metallic bulk $Cu$ and $Ag$, if one ignores the energy barriers needed to actually break all bonds in the nanoalloy.

On the experimental side, in addition, we note that the pure clusters $Ag_{7}^-$ (Ref. 47) and $Ag_{19}^+$ (Ref.~\cite{a48}) were found to have dissociation energy of 2.73 eV and 2.88 eV, which are very close to the experimentally observed cohesive energy of bulk $Ag$ (2.94 eV). The formation of pure cluster structures, such as $Ag_{7}^-$ and $Ag_{19}^+$, may thus be seen to be energetically more favorable than the nanoalloys. Perhaps, the possible disintegration of $Ag_{27}Cu_{7}$ into pure-element clusters may be argued against on the basis of the strength of the $Ag-Cu$ bond. To estimate the strength of the bonds in the 34-atom nanoalloys and understand what distinguishes $Ag_{27}Cu_{7}$ in its family of nanoalloys, we turn to Fig.2 of Ref.~\cite{a1}.  Rossi et al. show that as the amount of $Cu$ increases up to $\sim$20 {\%} (starting from $Ag_{34}$), the heat of formation is reduced or kept constant, implying that small amounts of $Cu$ atoms immersed among $Ag$ atoms (in the nanoalloys $Ag_{34-n}Cu_{n}$ as $n$ decreases from 7 to 1) create $Cu-Ag$ and $Cu-Cu$ bonds that are stronger than those in bulk $Cu$ and are able not only to counterbalance the cost of the cohesive energy of the newly added $Cu$ atom, but also to increasingly stabilize the nanoalloy. Fig.2 of Ref.~\cite{a1} shows also that if the content of $Cu$ increases beyond 7 atoms, the heat of formation increases again, with the implication that the strength of the bonds is not able to compensate the bulk $Cu$ cohesive energy for an additional atom. In conclusion, $Ag_{27}Cu_{7}$ possesses the composition and geometry that maximizes the number of $Cu-Cu$ and $Cu-Ag$ bonds using the minimum number of $Cu$ atoms.

The structural stability considerations presented above are derived entirely from the energetics of $Ag_{27}Cu_{7}$. The contributions of vibrational entropy could be important~\cite{a18} and may lead to a minimum of the free energy (as a function of the $Ag/Cu$ ratio) which is different from that in the plot of Rossi et al. They, however, argue against such a possibility.~\cite{a1} Very recent calculations~\cite{TSR1} based on many-body interaction potentials also confirm that vibrational entropic contributions are small for this set of 34-atom nanoalloys. Calculations of the phonon frequencies of these nanoalloys from first principles, as presented in section A.3 for the $Ll_{2}$  bulk alloys, are still desirable, as such a study could serve as an indicator of stable compositions. However, the resulting contributions of the vibrational entropy to the free energy are expected to be small and should not change the conclusions drawn here. Configurational entropy may in general also play a role in determining the stable configurations of these nanoalloys, but it is not expected to be important for $Ag_{27}Cu_7$ because of the lack of degenerate isomers.~\cite{TSR2}

\subsection{Electronic structure and charge density distribution}\label{acIIIB}

\subsubsection{Electronic DOS of $Ag_{27}Cu_{7}$ nanoalloy and $Ll_{2}$  $Ag-Cu$ alloys}\label{acIIIB1}

In our calculations the $Ag_{27}Cu_{7}$ nanoalloy has 374 valence electrons occupying 187 discrete, flat, and double degenerated bands, as shown in Fig.~\ref{fig:ac15}. As we can see, the discrete $d$-bands are densely concentrated between -5.0 and -1.5 eV below the Fermi level. Such an ideal behavior of the DOS may not be observed in experiments.~\cite{Issendorf} One of the reasons is the Jahn-Teller deformation which lifts the degeneracy of the levels, smears out the ideal DOS, and may also lead to the opening of gaps not predicted by the present approach.~\cite{Issendorf} The second reason is the actual atomic positions occurring in a cluster due to deviations from the ideal structure further modifies the DOS.~\cite{Issendorf} For example, Ref.~\onlinecite{Issendorf} shows that the photo-electron spectrum of $Cu_{34}^-$ (among others smaller and larger) is significantly smoother than 187 sharp peaks that one may expect, though it does displays smooth dips. To obtain the electronic DOS of bulk $Cu_{3}Ag$ and $Ag_{3}Cu$  alloys and $Ag_{27}Cu_{7}$ nanoalloy from \emph{ab initio} electronic structure calculations, the states are thus broadened using Gaussian functions of width 0.14 eV.

Our resulting DOS of $Ag_{27}Cu_{7}$ is shown in Fig.~\ref{fig:ac5}(a)-(c), while that of the bulk alloys is presented in Fig.~\ref{fig:ac5}(d) and (e). First of all, the HOMO-LUMO gap ($\Delta$ in Fig.~\ref{fig:ac5}(a)) of $Ag_{27}Cu_{7}$ is found to be 0.77 eV in the ground state, which is only slightly smaller than that reported by Rossi et al,~\cite{a1}
0.82 eV. As expected, the $s$-states have negligible contributions and the displayed structures have mostly $d$-character between -5.3 and -1.5 eV. Fig.~\ref{fig:ac5}(b) shows that the centroid ($\frac{\int E  f(E) dE}{\int f(E) dE}$) of the nanoalloy valence band red-shifts $\sim$~1 eV as compared to bulk $Ag$, and blue-shifts $\sim$~0.5 eV with respect to bulk $Cu$. Fig.~\ref{fig:ac5}(c) shows that even though the amount of $Cu$ in the $Ag_{27}Cu_{7}$ nanoalloy is $\sim$~4 times less than that of $Ag$, it contributes to blue-shift the valence band. As shown in Fig.~\ref{fig:ac5}(c)-(e) and in Refs.~\cite{a17,a49}, the role of $Cu$ is, in general, to enhance the DOS at the top of the valence band and to blue-shift the centroid, while the effect of $Ag$ is the opposite. Similar results have been reported for
$Au-Pd$ nanoclusters,~\cite{a52} in which the increasing content of $Au$ on $Pd$ clusters reduces the density of states at the Fermi level. Fig.~\ref{fig:ac5}(b) and (c) show in addition that the $d$-band of the nanoalloy is as broad as that of either pure bulk constituents - a point worth of noticing since atoms in a low coordinated environment generally exhibit a valence band narrowing.~\cite{a50} Pure $Ag$ clusters and $Cu_{3}Au$  surfaces,~\cite{a49, a50} for example, have shown this effect. The hybridization of $Ag$ and $Cu$ states in $Ag_{27}Cu_{7}$ thus compensates the $d$-band narrowing that each atom undergoes.

In general, the features in the electronic DOS that discriminate stable alloy phases are not yet fully understood. Although it is well accepted~\cite{a34}  that the dip in the DOS at the Fermi level is related to the stability of a particular alloy phase, there is no obvious correlation between the two since stable and ordered $Cu-Au$ alloys present a dip in the
DOS while $Au-Ag$ alloys do not.~\cite{a34}  In the former the dip in the DOS changes position, width and depth with composition, and structure, and has been found to be related to the electronic specific heat.~\cite{a17} Kokko et al.~\cite{a17} noticed also that the dip is considerably lessened in the layered (tetragonal $L1_0$) $CuAu I$ phase (which reduces the $Au-Cu$ bond) with respect to $Cu_{3}Au$  and $Au_{3}Cu$. Also, based on their electronic specific heat calculations, they infer that the dip is even smoother in the disordered phases. In this work, we find that the nanoalloy $Ag_{27}Cu_{7}$ displays a
less pronounced dip (deepest at $\sim$ - 2.25 eV) in the $d$-band region (between $\sim$ -5.3 and $\sim$ -1.5 eV) - where $Cu$ and $Ag$ states hybridize - than those displayed by $Ll_{2}$  $Cu_{3}Ag$ and $Ag_{3}Cu$ alloys (deepest at $\sim$ -3.6 eV and $\sim$ -2.5 eV, respectively. Compare Fig.~\ref{fig:ac5}(a), (d) and (e)), which are  similar to, but broader than, those found in
stable $Cu-Au$ alloys.~\cite{a17} Yet, from considerations of the heat of formation, the $Ag-Cu$ alloys are marked as being immiscible. Furthermore, we find that the DOS of the $Ag$ atoms (Fig.~\ref{fig:ac5}(b) and (c)) in $Ag_{27}Cu_{7}$ resembles to some extent that of bulk $Ag$, despite being highly under coordinated, while that of the fully coordinated $Cu$ atoms is strikingly different from bulk $Cu$ presumably, since half of its neighbors are $Ag$ atoms.

In order to understand the correlation between the $Ag/Cu$ content ratio, the consequent decrease/increase of the bond lengths, and the structure of the electronic DOS, we turn to examination of the changes experienced by the DOS of individual $Ag$ and $Cu$ atoms in a set of environments (some natural, some artificial): $Ag_{27}Cu_{7}$ nanoalloy, a
free standing $Ag$ and $Cu$ monolayer, bulk $Ag-Cu$ alloys, and bulk $Ag$ and $Cu$ (Figs.~\ref{fig:ac6}-\ref{fig:ac8}).  For the bulk systems, we consider also the effect on the DOS (Fig.~\ref{fig:ac7} of expanding and contracting the lattice constant from the equilibrium value found in our DFT calculations. The DOS of $Ag$ and $Cu$ atoms in
Figs.~\ref{fig:ac6}-\ref{fig:ac8} allow comparison on a \emph{one-to-one} basis and not as percentile contributions as presented in Fig.~\ref{fig:ac5}.

From Figs.~\ref{fig:ac6}(a) and (b), we note that decrease in the content of $Ag$ (and thus the lattice parameter) gives rise to a sharp peak at intermediate energies while the bottom of the d-band remains almost unchanged and states from the top of the bulk $Ag$ band (Fig.~\ref{fig:ac6}(e)) retract to lower energies. A few states, however, appear between -1.5 and -2.5 eV, above the range of the DOS of bulk $Ag$, and hybridize with $Cu$ states. As we shall see, the appearance of these higher energy states is related purely to the presence of $Cu$, whereas the depletion of the top of the $Ag$ $d$-band results from both the presence of $Cu$ and from the enhanced overlap of $Ag-Ag$ orbitals as a result of the decrease in the $Ag-Ag$ bond length. 

To isolate the effect of the bond length, we turn to the DOS in Fig.~\ref{fig:ac7} for $Ag$ and $Cu$ atoms in bulk environments with bond lengths different from the equilibrium values. In order to maintain a reference point, we have taken the lattice constant of expanded $Ag_3Cu$ and $Cu_3Ag$ to be that of bulk $Ag$, corresponding to an expansion of 2.4~{\%} in the former and 8.9{~\%} in the latter. Similarly, to infer the effect of lattice contraction we have used the lattice constant of bulk $Cu$ for $Ag_3Cu$ (contraction of 9.8{~\%}) and the bond-length of $Cu0-Cu1$ for $Cu_3Ag$ (contraction of 5.5{~\%}). By comparing Figs.~\ref{fig:ac6}(e) and \ref{fig:ac7}(f), we find that contraction of $Ag-Ag$ bonds pushes the bottom of the $Ag$ $d$-band to lower energies and depletes the top of the band, while comparison of Figs.~\ref{fig:ac6}(a) with \ref{fig:ac7}(b) and \ref{fig:ac6}(b) with \ref{fig:ac7}(d) indicates that the contraction of $Ag-Cu$ bonds also pushes the bottom of the $Ag$ $d$-band to lower energies and significantly lessens the highest features of the DOS of $Ag$ $d$-band in bulk alloys. We also conclude from Figs.~\ref{fig:ac7}(b) and \ref{fig:ac7}(d), and Fig.~\ref{fig:ac6}(e), \ref{fig:ac6}(a) and \ref{fig:ac6}(b) that the presence of $Cu$ is responsible for the appearance of $Ag$ states above the top of the bulk $Ag$ band hence improving the overlap with the $Cu$ $d$-band. On the other hand, as seen from Figs.~\ref{fig:ac6}(e) and \ref{fig:ac7}(e), the expansion of $Ag-Ag$ bonds depletes the bottom of the DOS and slightly enhances the DOS at the top. Similarly, the expansion of the $Ag-Cu$ in the bulk alloys depletes the bottom of the DOS and enhances the highest peaks of the DOS (compare \ref{fig:ac6}(a) with \ref{fig:ac7}(a) and \ref{fig:ac6}(b) with \ref{fig:ac7}(c)).  
Atomic low coordination and expansion of the bond length cause the same effects on the DOS of $Ag$, though these are augmented in the former situation. Such was found to be the case of a free standing $(111)$ monolayer (Fig.~\ref{fig:ac6}(d)), in which the 6-coordination of $Ag$ atoms causes a strong depletion of the bottom and enhancement at the top of the $d$-band. 

From the above, we conclude that the low coordination of $Ag$ atoms in the $Ag_{27}Cu_{7}$ nanoalloy can indeed account for the depletion of the bottom of the $d$-band (Fig.~\ref{fig:ac6}(c)). On the other hand, the reduction of the DOS at the top of the $d$-band of $Ag0$ (Fig.~\ref{fig:ac8}(a)), $Ag3$ (Fig.~\ref{fig:ac8}(c)), and $Ag4$ (Fig.~\ref{fig:ac8}(d)) atoms, suggests that the presence of $Cu$ at such short distances outweighs the effect of their low coordination and the DOS at the top of the $Ag$ $d$-band is not enhanced.
The DOS of $Ag2$ is slightly more complex. As mentioned in subsection~\ref{acIIIA2}, $Ag2$ is the most undercoordinated atom of the cluster (See Fig.~\ref{fig:ac3}). Consistently, it has the least number of states below -4.5 eV and highest DOS above $\sim$ -2.8 eV, implying that low coordination effects on the DOS of $Ag2$ are not completely washed out through the short bond with $Cu0$, each one of which is incidentally shared by two $Ag2$ atoms. At the same time, as seen in bulk alloys, the hybridization of $Cu$ and $Ag$ states is improved by setting off the occupation of states above the top of the $d$-band of bulk $Ag$ (see Fig.~\ref{fig:ac6}(c) and Figs.~\ref{fig:ac8}(a)-(d)).

We now turn to the issue of the DOS of $Cu$ atoms in the $Ag_{27}Cu_{7}$ nanoalloy which differ prominently from that of bulk $Cu$. We consider three different aspects that may influence their electronic structure: the conspicuously disparate overall geometry, the presence of $Ag$, and the existence of bond lengths longer than that of bulk $Cu$. To understand each one of these we first note that if $Cu-Ag$ bonds are longer than that of bulk $Cu$, then the bottom of the $d$-band of $Cu$ is strongly depleted and an increasingly sharper peak at the top is created, which slightly shifts towards higher energies (compare Figs.~\ref{fig:ac6}(f) with \ref{fig:ac7}(g) and \ref{fig:ac6}(g) with \ref{fig:ac7}(i)). As already noted for the Ag atoms, the effect of expanding $Cu-Cu$ bonds is similar to that of $Cu$ atoms in a low coordinated environment (compare Fig.~\ref{fig:ac7}(k) and \ref{fig:ac6}(i) with \ref{fig:ac6}(j)), that is, the bottom of the $Cu$ $d$-band (states below $\sim$3.5 and $\sim$~4.0 eV in Figs.~\ref{fig:ac6}(i) and \ref{fig:ac7}(k), respectively) is entirely extinguished. Interestingly, the $d$-band of the free monolayer is sharply localized at the top edge, resembling that of $Cu0$ (Fig.~\ref{fig:ac8}(e)) and $Cu1$ (Fig.~\ref{fig:ac8}(f)) atoms notwithstanding their full coordination in $Ag_{27}Cu_{7}$. Contracting $Ag-Cu$ bonds in bulk alloys reduces the DOS at the top, which is pushed to lower energies, and enhances the bottom of the $d$-band, thus improving the $Ag-Cu$ hybridization, as seen by comparing Figs.~\ref{fig:ac6}(f) with \ref{fig:ac7}(b), \ref{fig:ac6}(g) and \ref{fig:ac7}(j). Similar features are found by contracting the $Cu-Cu$ bond is bulk $Cu$ (compare Figs.~\ref{fig:ac6}(j) and \ref{fig:ac7}(l)). More importantly, note in Figs.~\ref{fig:ac7}(j) that even when $Cu-Cu$ and $Cu-Ag$ bonds in bulk $Cu_{3}Ag$  are shorter than that of bulk $Cu$, the region from $\sim$~4.0 eV . $\sim$~5.0 eV is nevertheless strongly quenched, suggesting that the presence of $Ag$ also strongly depletes most of the bottom of the $d$-band of $Cu$ atoms (something similar occurs in $Au-Cu$, see Ref.~\cite{a17}). Observe also that presence of $Ag$ introduces states below the bottom of the DOS of bulk Cu (Figs.~\ref{fig:ac6}(f)-(h) and \ref{fig:ac7}(g)-(j)), which hybridize with $Ag$. 

From the above, we conclude first that the sharp peak (Fig.~\ref{fig:ac6}(h)) at the top of the DOS of $Cu$ atoms in $Ag_{27}Cu_{7}$ nanoalloy, characteristic of low-coordination (Figs.~\ref{fig:ac6}(i)), can only be accounted for by the relatively long distances ($\sim$~2.7-2.8 {\AA}) between $Cu$ atoms and half of their \emph{nearest neighbors} - all $Ag$ atoms - (see Figs.~\ref{fig:ac3}(a) and Table I), seemingly leading to a weak interaction of the $Cu$ atoms with those \emph{far-lying} $Ag$ neighbors, as occurs for $Cu$ atoms in bulk $Ag_{3}Cu$  and $Cu_{3}Ag$  alloys. The DOS of $Cu$ atoms in bulk $Ag_{3}Cu$  (Figs.~\ref{fig:ac6}(f)) thus indicates that the strength of the $Ag-Cu$ bond is considerably weak for $Cu$ atoms; in fact, expanding the lattice parameter (Figs.~\ref{fig:ac7}(g)) changes insignificantly the DOS of $Cu$. Second, the effect of the $Ag$ environment on the DOS of $Cu$ atoms in $Ag_{27}Cu_{7}$ is connected with the depletion of states between 2.5 and 3.5 below the Fermi level, i.e., the dip around 3.0 eV in Fig~\ref{fig:ac6}(h), \ref{fig:ac6}(e) and (f). We note in addition that the DOS of $Ag_{27}Cu_{7}$ below the dip ($\sim$~3.5 eV) is remarkably high, as compared to that of bulk alloys (see Figs.~\ref{fig:ac6}(f) and (g)) - despite the low $Cu$ content - and generates a much stronger hybridization between $Cu$ and $Ag$ states, contrasting that observed in bulk alloys (see Figs.~\ref{fig:ac5}(c)-(e)); interestingly, the DOS of states of $Cu$ atoms at 3.5 eV below the Fermi level in the compressed lattice of bulk $Ag_{3}Cu$  (Figs.~\ref{fig:ac7}(b)) is almost as high as that of $Cu$ atoms in $Ag_{27}Cu_{7}$ and also results in a stronger hybridization between $Cu$ and $Ag$ states compared to that found in equilibrium bulk $Ag_{3}Cu$  bulk. In the third place, we conclude that the \emph{optimum} $Cu-Ag$ bonding in $Ag_{27}Cu_{7}$ comes about in terms of the electronic DOS through the shortening of $Ag-Cu$ bonds that allows the hybridization of the $d$-states of $Cu$ and $Ag$ atoms, suppressing by this means the dip featuring in the DOS of $Cu_{3}Ag$  and $Ag_{3}Cu$  alloys.

Finally, we remark that although the DOS of $Ag$ atoms is not changed as drastically as that of $Cu$ atoms in bulk $Ag-Cu$ alloys - suggesting that $Cu$ is more sensitive than $Ag$ to the chemical environment -, the vulnerability of $Cu$ to the presence of $Ag$ is intermixed with long bond-length effects. That is, the presence of $Cu$ depletes the top of the $Ag$ $d$-band as much as $Ag$ depletes the bottom of $Cu$ $d$-band. However, the short $Ag-Ag$ and $Ag-Cu$ bonds (with respect to bulk $Ag$) as present in both bulk alloys broaden the $d$-band of $Ag$ atoms, compensating in this manner the effect of the $Cu$ atoms. On the other hand, the $Cu-Cu$ and $Cu-Ag$ bond lengths induced in both bulk alloys never become shorter than that of bulk $Cu$, rather it is quite the opposite. As a result, narrowing of the $d$-band of $Cu$ atoms is triggered, aggravating the depletion caused by the presence of $Ag$ atoms and, hence, exaggerating the actual chemical susceptibility of $Cu$ to the $Ag$ environment.
In summary, the electronic DOS is found unambiguously related to the bond lengths held in a particular geometry. 

If we were to extrapolate the above results to related systems, we would speculate  that the dip in the DOS identifies less stable phases of noble metal alloys. For example, based on the patterns found in $Ag_{27}Cu_{7}$ and the $L1_2$ bulk $Ag-Cu$ alloys, we predict that the stability $Au-Ag$ alloys and the absence of a dip in their DOS~\cite{a34} is attributed to the fact that the lattice parameter of bulk $Au$ and $Ag$ are nearly identical and that the $d$-band of bulk $Ag$ lies within that of bulk $Au$, assuring significant $d$-band hybridization. Likewise, the hybridization between $Cu$ and $Au$ states is strong in bulk $Au-Cu$ alloys because the $d$-band of $Cu$ also lies within that of bulk $Au$, albeit in the region near its Fermi level, while that of $Ag$ lies deeper. The overlap between the bands of $Ag$ and $Cu$, on the other hand, is relatively small resulting in weaker $Cu-Ag$ hybridization in bulk $Cu-Ag$ alloys. The dip in the DOS of $Au-Cu$ is probably not a sign of stability, rather it may be a sign that structures and/or compositions allowing shorter $Au-Cu$ would be more stable (amorphous phases, perhaps). The heat of formation of bulk $Cu_{3}Au$, for example, is negative because the strength of the $Au-Cu$ bond is larger than that of bulk $Cu$ and $Au$, even at the distances dictated by the $L1_2$ phases,~\cite{naval} but which are not necessarily the distances at which the strength of all three $Au-Cu$, $Au-Au$, and $Cu-Cu$ bonds is optimized with the corresponding hierarchal importance, as exemplified by $L1_{2}$  bulk $Ag-Cu$ alloys whose long $Ag-Cu$ bonds contrast the significantly short ones found in the $Ag_{27}Cu_{7}$ nanoalloy.

\subsubsection{Charge density distribution of $Ag_{27}Cu_{7}$ nanoalloy}\label{acIIIB2}

The first aspect that comes to mind in Fig.~\ref{fig:ac9}-\ref{fig:ac14} of the plotted charge densities of the nanoalloy $Ag_{27}Cu_7$ is that $Ag$ atoms barely supply charge to the surface of the nanoalloy. The question is whether the surface charge depletion coincides with  charge redistribution from $Ag$ atoms to $Cu$ atoms since two neighboring metals with significantly different work function can give rise to electron transfer from one metal to the other, as reported in calculations of $Pd$ clusters on $Au(111)$ by S{\'a}nchez et al,~\cite{a52}  in correspondence with a work function difference of $\sim$~0.3 eV. Here, the work function of $Cu$ is larger
than that of $Ag$ by $\sim$~0.2 eV. Indeed, Figs.~\ref{fig:ac9} and \ref{fig:ac10} display higher charge density around $Cu$ atoms than around $Ag$ atoms.

The plot of the charge density in a plane passing through 10 $Ag$ and $Cu$ atoms in Fig.~\ref{fig:ac9} illustrates that the $Cu0$-$Ag2$ and $Cu1$-$Ag4$ bonds are linked by the highest bonding charge density, corresponding indeed to the shortest $Ag-Cu$ bond lengths ($\sim$~2.58 {\AA}), and implying that they are stronger than the $Cu1$-$Cu1$ (~2.58 {\AA}), $Cu0$-$Cu0$ (2.58 {\AA}) bonds and even the $Cu0$-$Cu1$ bonds (2.55 {\AA}), in that order (see Figs.~\ref{fig:ac9} and \ref{fig:ac10}). The next strength of bonding charge density occurs for $Cu0$-$Ag0$ bonds, followed by $Cu1$-$Ag3$ bonds, as
shown in Figs.~\ref{fig:ac9} and \ref{fig:ac10}. In this case, the bond lengths are 2.72 and 2.73 {\AA}, respectively. Notice that, in the second set of bonds, the charge density is considerably lower than in $Cu0$-$Ag2$ and $Cu1$-$Ag4$ bonds, which supports the assumption that $Ag-Cu$ interactions die out very rapidly (see Section~\ref{acIIIB1}). The third place in bonding charge density corresponds to $Cu0$-$Ag3$ (Fig.~\ref{fig:ac11}) bonds and the shortest $Ag-Ag$ bonds: $Ag0$-$Ag3$ and $Ag2$-$Ag2$ bonds (Fig.~\ref{fig:ac9}) whose bond lengths are 2.79, 2.87 and 2.89 {\AA}, respectively (Table~\ref{tab:table1}). The first might influence very little $Cu0$ atoms since the bond is quite large. However, this bonding charge density appears as large as that of $Ag0$-$Ag3$, which is 0.1 {\AA} further apart. The next in bonding charge density is the $Ag2$-$Ag3$ bond (Fig.~\ref{fig:ac12}), whose length is 2.90~{\AA}, followed by the $Ag3$-$Ag4$
bond (Fig.~\ref{fig:ac9}), whose length is 2.93 {\AA}. The latter bond length is close to that of the $Ag0$-$Ag3$ bond (and almost identical to the bond length of bulk $Ag$); however, the charge density bridging these atoms is slightly weaker. Note that $Ag3$-$Ag3$ bonds (Fig.~\ref{fig:ac13}) are only slightly longer (2.95 {\AA}) but the bonding charge density around it is less than that around the $Ag3$-$Ag4$ bonds.  The charge
density around the $Ag0$-$Ag2$ bond (Fig.~\ref{fig:ac14}) is substantially lower than that around $Ag3$-$Ag3$, consistent with a larger bond length, 3.0 {\AA}. The next larger bond lengths are more than $\sim$~4 {\AA}, which are expected to provide much weaker direct interactions that will not be discussed here.

In reference to the importance of bond strength and length hierarchies in alloys, mentioned in subsection~\ref{acIIIB1}, we turn to the discussion in~\ref{acIIIA3} of Ref.~\cite{a38} in which $Ag-Ru$ phonon-stable structures were associated with a high charge density bridging atoms of the same element, whereas phonon-unstable structures were associated with a high charge density bridging atoms of different elements. Instead, following the arguments above, we propose that their
charge density plots indicate that $Ag$-rich structures do not allow for strong $Ru-Ru$ bonds and so, although strong $Ru-Ag$ and $Ag-Ag$ bonds are present, the structure is unstable. Additional support to our assertion comes from the fact that phonon instabilities are not present in some $Ru$-rich structures,~\cite{a38} while those in $Ag$-rich structures disappear by reducing the equilibrium lattice parameter.~\cite{a38} Besides, $Ag$-rich structures with reduced lattice parameter show clearly that the charge density bridging $Ru-Ru$ bonds is enhanced (see Fig.~\ref{fig:ac3}(e) of Ref.~\cite{a38}). The above suggests that in $Ag-Ru$ alloys the strength hierarchy is in the following order: \{$Ru-Ru$, $Ag-Ru$, $Ag-Ag$\}. In fact, relatively large differences between the strength of $Ru-Ru$ and $Ag-Ru$ bonds and a pronounced deep dip in the DOS can be expected to explain all their results simultaneously.

\section{SUMMARY}\label{acIV}

We have presented DFT calculations of the $Ag_{27}Cu_{7}$ nanoalloy to understand its structure and relative stability via considerations of its energetics, electronic DOS, and charge density distribution. The local coordination of $Cu$ atoms is similar to that in bulk $Cu$ regarding the number of first and second NN, whereas $Ag$ atoms find themselves in a  low-coordinated environment but, in exchange, form $Ag-Cu$ bonds which are short as the shortest $Cu-Cu$ bonds. On the other hand, the electronic structure of the $Cu$ atoms in $Ag_{27}Cu_{7}$ deviates much more from that of atoms in bulk $Cu$, as compared to the corresponding case of $Ag$ atoms in this nanoalloy.

Related bulk alloys, $Ag_{3}Cu$  and $Cu_{3}Ag$, are found to have positive heat of formation and form larger bonds than the shortest ones found in the $Ag_{27}Cu_{7}$. However, we find that the resulting interatomic bonds in these bulk alloys are sufficiently strong such that their cohesive energy is larger than that of bulk $Ag$ and their phonon dispersion curves do not display instabilities.

From our analysis of the geometric and electronic structure, we conclude that the relative stability of $Ag_{27}Cu_{7}$, among its nanoalloy family, is the result of the maximization of the number of $Cu-Cu$ and $Cu-Ag$ bonds, using the minimum number of $Cu$ atoms. The core-shell $Ag-Cu$ nanoalloys do not behave differently from the corresponding bulk alloys regarding segregation tendencies and migration of $Ag$ to the surface, as pointed earlier,~\cite{a37} since $Ag_{27}Cu_{7}$ is segregated by construction. Furthermore, the segregated structure is the attribute that leads to its relative stability, provided  the core-shell structure allows formation of strong $Cu-Cu$ and $Cu-Ag$ bonds without contracting the typical $Ag-Ag$ bond or (most importantly) stretching the typical $Cu-Cu$ bond.

The HOMO-LUMO gap of $Ag_{27}Cu_{7}$ is found to be 0.77 eV. The DOS of $Ag_{27}Cu_{7}$ shows  features similar to those of $Ag-Cu$ bulk alloys. We find that the novel  features of the DOS of fully coordinated $Cu$ atoms in $Ag_{27}Cu_{7}$ are caused by the relatively long distance separating Cu atoms from half of its first NN. Short $Ag-Cu$ bond-lengths, on the other hand, improve the hybridization of $Cu$ and $Ag$ states; explaining why the hybridization of $Cu$ and $Ag$ states in the $Ag_{27}Cu_{7}$ nanoalloy is stronger than in $Ag-Cu$ bulk alloys. The observed differences in electronic DOS between $Ag_{27}Cu_{7}$ and $L1_2$ alloys arise not only because of low-coordination and geometry differences but mainly because the symmetry enforces long $Cu-Cu$ and $Cu-Ag$ bonds, unlike the situation in $Ag_{27}Cu_{7}$.

In $Ag_{27}Cu_{7}$, the charge density along $Ag-Cu$ bonds whose length is $\sim$~2.6 {\AA} is even larger than that around $Cu-Cu$ bonds, and certainly larger than that bridging $Ag-Ag$ atoms. Nevertheless, $Ag-Cu$ bonds whose length is of the order of that in bulk $Ag_3Cu$, or even $Cu_{3}Ag$, are surrounded by an appreciably quenched charge density, explaining why the DOS of $Cu$ atoms show low coordination features. 

We infer a hierarchy of bond strength intrinsic to $Cu$ and $Ag$: $\{Cu-Cu > Cu-Ag > Ag-Ag\}$. The nanoalloy structures that enforce such a hierarchy also obey a bond-length order, $\{ Ag-Ag > Cu-Ag > Cu-Cu \}$. Clearly, the strength and length of the bonds are correlated properties in any particular structure. Hence, for a given  composition of a nanoalloy, the global-minimum structure will be that which simultaneously maximizes the number of the strongest bonds while maintaining the bond-length order, and thus fulfilling the bond strength hierarchy that ultimately  minimizes the energy. 

%We infer a hierarchy of bond strength, \{$Cu-Cu$, $Cu-Ag$, and $Ag-Ag$\}, correlated to a bond length order, \{$Ag-Ag$, $Cu-Ag$, and $Cu-Cu$\} so that, the actual strength of the bonds in a particular structure becomes the signature of a relaxed geometry, which may or may not satisfy the above hierarchy, let the strongest bonds to realize, and thus be stable.

\begin{acknowledgments}
We are indebted to Prof. Riccardo Ferrando for providing us with the initial configuration of the $Ag_{27}Cu_7$ nanoalloy. This work was supported in part by DOE under grant DE-FG02-03ER46058.
%\dots.
\end{acknowledgments}

\clearpage

\begin{table*}
\caption[$Ag_{27}Cu_7$ nanoalloy bond lengths]{\label{tab:table1}
This table contains six sets of 2-column \emph{sub-tables} showing the distance from each type of atom in the nanoalloy to all its neighbors. In the right column of each sub-table appears the type of neighbor which is being referred and the number of such equivalent atoms at the same distance is shown in parenthesis. Notice that here equivalent atoms are not considered being so if they do not belong to the same layer.
}
\begin{ruledtabular}
\begin{tabular}{llllllllllll}
\multicolumn{2}{c}{$Cu0$}&\multicolumn{2}{c}{$Ag0$}&\multicolumn{2}{c}{$Cu1$}&\multicolumn{2}{c}{$Ag2$}&\multicolumn{2}{c}{$Ag3$}&\multicolumn{2}{c}{$Ag4$} \\
%\multicolumn{2}{|c|}{$Cu0$}&\multicolumn{2}{|c|}{$Ag0$}&\multicolumn{2}{|c|}{$Cu1$}&\multicolumn{2}{|c|}{$Ag2$}&\multicolumn{2}{|c|}{$Ag3$}&\multicolumn{2}{|c|}{$Ag4$} \\
\hline
Type&&Type&&Type&&Type&&Type&&Type& \\
(NN)&$d$(\AA)&(NN)&$d$(\AA)&(NN)&$d$(\AA)&(NN)&$d$(\AA)&(NN)&$d$(\AA)&(NN)&$d$(\AA)\\
\hline
Cu1$^{(1)}$& 2.548   &Cu0$^{(2)}$&        2.715   &Cu0$^{(5)}$&        2.548   &Cu0$^{(1)}$&        2.587   &Cu1$^{(1)}$&        2.726   &Cu1$^{(1)}$&        2.590\\
Cu-1$^{(1)}$&        2.548   &Ag3$^{(1)}$&        2.882   &Cu-1$^{(1)}$&       2.576   &Ag-2$^{(1)}$&       2.867   &Cu0$^{(2)}$&        2.786   &Ag3$^{(5)}$&        2.932\\
Cu0$^{(2)}$& 2.584   &Ag-3$^{(1)}$&       2.882   &Ag4$^{(1)}$&        2.590   &Ag3$^{(2)}$&        2.902   &Ag0$^{(1)}$&        2.882   &Cu0$^{(5)}$&        4.457\\
Ag2$^{(1)}$& 2.587   &Ag2$^{(2)}$&        3.002   &Ag3$^{(5)}$&        2.726   &Ag0$^{(2)}$&        3.002   &Ag2$^{(2)}$&        2.902   &Ag2$^{(5)}$&        4.991\\
Ag-2$^{(1)}$& 2.587   &Ag-2$^{(2)}$&       3.002   &Ag2$^{(5)}$&        4.354   &Cu1$^{(1)}$&        4.354   &Ag4$^{(1)}$&        2.932   &Cu-1$^{(1)}$&       5.166\\
Ag0$^{(2)}$&  2.715   &Cu1$^{(1)}$&        4.360   &Ag0$^{(5)}$&        4.360   &Cu0$^{(2)}$&        4.462   &Ag3$^{(2)}$&        2.948   &Ag0$^{(5)}$&        5.691\\
Ag-3$^{(2)}$&        2.786   &Cu-1$^{(1)}$&       4.360   &Ag-3$^{(5)}$&       4.425   &Ag-3$^{(2)}$&       4.684   &Cu-1$^{(1)}$&       4.425   &Ag-3$^{(5)}$&       6.721\\
Ag3$^{(2)}$& 2.786   &Ag3$^{(2)}$&        4.769   &Ag-2$^{(5)}$&       5.132   &Ag4$^{(1)}$&        4.991   &Cu0$^{(2)}$&        4.482   &Ag-2$^{(5)}$&       6.866\\
Cu0$^{(2)}$&         4.181   &Ag-3$^{(2)}$&       4.769   &Ag-4$^{(1)}$&       5.166   &Ag2$^{(2)}$&        5.115   &Ag-2$^{(2)}$&       4.684   &Ag4$^{(1)}$&        7.756\\
Ag4$^{(1)}$&         4.457   &Ag0$^{(2)}$&        4.897   &       &       &Cu-1$^{(1)}$&       5.132   &Ag-3$^{(1)}$&       4.716   &       &\\
Ag-4$^{(1)}$&        4.457   &Cu0$^{(2)}$&        5.277   &       &       &Ag3$^{(2)}$&        5.729   &Ag0$^{(2)}$&        4.769   &       &\\
Ag2$^{(2)}$& 4.462   &Ag4$^{(1)}$&        5.691   &       &       &Ag-2$^{(2)}$&       5.864   &Ag3$^{(2)}$&        4.770   &       &\\
Ag-2$^{(2)}$&        4.462   &Ag-4$^{(1)}$&       5.691   &       &       &Cu0$^{(2)}$&        6.426   &Cu0$^{(1)}$&        5.263   &       &\\
Ag3$^{(2)}$& 4.482   &Cu0$^{(1)}$&        6.364   &       &       &Ag-3$^{(2)}$&       6.807   &Ag-3$^{(2)}$&       5.561   &       &\\
Ag-3$^{(2)}$&        4.482   &Ag3$^{(2)}$&        6.790   &       &       &Ag-4$^{(1)}$&       6.866   &Ag2$^{(2)}$&        5.729   &       &\\
Ag3$^{(1)}$& 5.263   &Ag-3$^{(2)}$&       6.790   &       &       &Ag3$^{(1)}$&        6.921   &Ag-3$^{(2)}$&       6.707   &       &\\
Ag-3$^{(1)}$&        5.263   &Ag2$^{(2)}$&        7.039   &       &       &Ag0$^{(2)}$&        7.039   &Ag-4$^{(1)}$&       6.721   &       &\\
Ag0$^{(2)}$& 5.277   &Ag-2$^{(2)}$&       7.039   &       &       &Ag-3$^{(1)}$&       7.837   &Ag0$^{(2)}$&        6.790   &       &\\
Ag0$^{(1)}$& 6.364   &Ag0$^{(2)}$&        7.923   &       &       &Ag2$^{(2)}$&        8.277   &Ag-2$^{(2)}$&       6.807   &       &\\
Ag2$^{(2)}$& 6.426   &Ag2$^{(1)}$&        8.637   &       &       &Ag0$^{(1)}$&        8.637   &Ag2$^{(1)}$&        6.921   &       &\\
Ag-2$^{(2)}$&        6.426   &Ag-2$^{(1)}$&       8.637   &       &       &Ag-2$^{(2)}$&       8.759   &Ag-2$^{(1)}$&       7.837   &       &\\
\end{tabular}
\end{ruledtabular}
\end{table*}

\clearpage

\begin{figure*}
\includegraphics[width=1.0\textwidth]{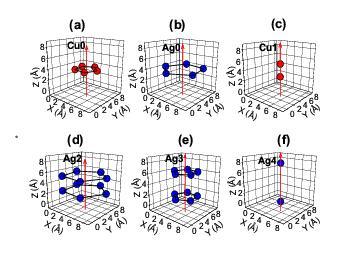}
\caption{\label{fig:ac1}
}
\end{figure*}

FIG.~\ref{fig:ac1} (Color online) Three-dimensional picture of the six types of non-equivalent atoms. The vertical arrow represents the 5-fold rotation axis. (a) Five $Cu0$ atoms sitting on the mirror plane of the cluster, z$_{0}$ = 3.88 {\AA}. Note that the z$_{i}$-coordinates are given with respect to the reference frame used in the figure and will define later interlayer distances; (b) The five $Ag0$ atoms also sit on the mirror plane, z$_{0}$ = 3.88 {\AA}; (c) Two $Cu1$ atoms sitting on the 5-fold rotation axis symmetrically located above and below the mirror plane, z$_{1}$= 5.17 {\AA}  and z$_{-1}$=2.59 {\AA}. Notice that $Cu0$-pentagon fits in the $Ag0$-pentagon; (d) Ten $Ag2$ atoms form two pentagons symmetrically located above and below the mirror plane, z$_{2}$ = 5.31 {\AA}  and z$_{-2}$ = 2.44 {\AA}; (e) Ten $Ag3$ form two pentagons symmetrically located above and below the mirror plane, z$_{3}$ = 6.24 {\AA} and   z$_{-3}$ = 1.52 {\AA}; (f) The two $Ag4$ atoms sit on the 5-fold rotation axis symmetrically located above and below the mirror plane, z$_{4}$ = 7.76 {\AA}  and z$_{-4}$ = 0 {\AA}.

\begin{figure}
\includegraphics[width=0.5\textwidth]{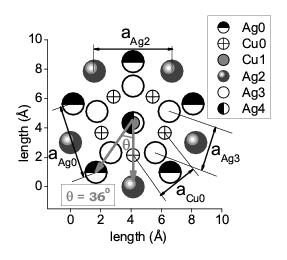}
\caption{\label{fig:ac2}
}
\end{figure}

FIG.~\ref{fig:ac2} \emph{Top view} of $Ag_{27}Cu_{7}$ nanoalloy, perpendicular to the mirror plane. The side length of each pentagon as $a_{M_{n}}$ (see text).

\begin{figure*}
\includegraphics[width=1.0\textwidth]{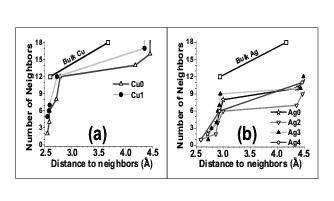}
\caption{\label{fig:ac3}
}
\end{figure*}

FIG.~\ref{fig:ac3} (a) Bond coordination for $Cu$ atoms of $Ag_{27}Cu_{7}$  compared with $Cu$ bulk; (b) bond coordination for $Ag$ atoms of $Ag_{27}Cu_{7}$  compared with $Ag$ bulk.

\begin{figure*}
\includegraphics[width=0.8\textwidth]{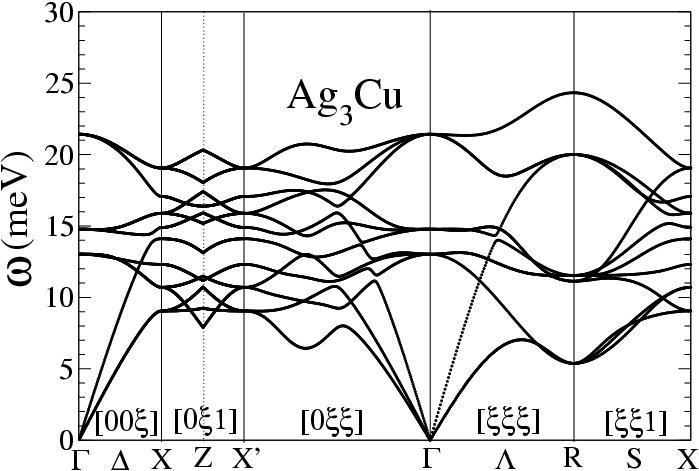}
\vskip 1.0in minus 1.9in
\includegraphics[width=0.8\textwidth]{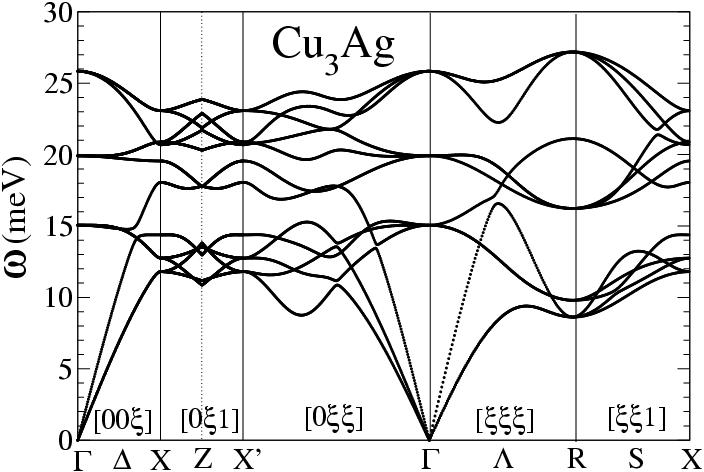}
\caption{\label{fig:ac4}
}
\end{figure*}

FIG.~\ref{fig:ac4} Calculated phonon spectra of the hypothetical $L1_{2}$  phase of (a) $Ag_{3}Cu$  and (b) $Cu_{3}Ag$ bulk alloys.

\begin{figure}
\includegraphics[width=0.5\textwidth]{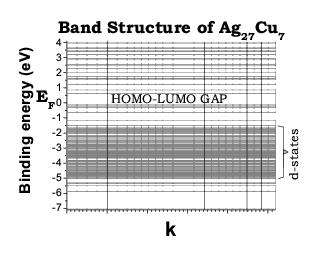}
\caption{\label{fig:ac15}
}
\end{figure}

FIG.~\ref{fig:ac15} Band structure of $Ag_{27}Cu_7$ showing that the bands do not depend on $\boldsymbol{k}$ (see Ref.~\onlinecite{comment}) and that the  $d$-bands are concentrated between -5.0 and -1.5 eV below the Fermi level (see text).

\begin{figure*}
\includegraphics[width=1.0\textwidth]{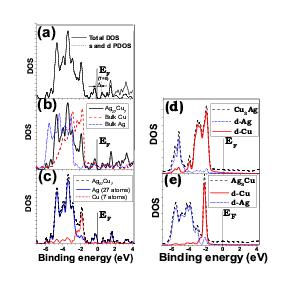}
\caption{\label{fig:ac5}
}
\end{figure*}

FIG.~\ref{fig:ac5} (Color online) (a) Total and projected electronic DOS of $Ag_{27}Cu_{7}$. The later corresponds to the contribution of $s$ and $d$ atomic states between -7 and 4.2 eV from the Fermi level, $E_F$, which is set equal to 0. The $s$-contribution is negligible up to $E_F$. The HOMO-LUMO gap in the ground state, $\Delta$ = 0.77 eV, is highlighted in red; (b) comparison between the total DOS of $Ag_{27}Cu_{7}$ and that of Ag and Cu bulk; (c) contribution from the core ($Cu$) and shell ($Ag$) atoms to the projected DOS of $Ag_{27}Cu_{7}$; (d) DOS of $Cu_{3}Ag$ showing the $d$-contribution from each species to the total DOS of the alloy in the $L1_{2}$ phase; (e) DOS of $Ag_{3}Cu$ showing the $d$-contribution from each species the total DOS of the alloy in the $L1_{2}$ phase.

\begin{figure*}
\includegraphics[width=1.0\textwidth]{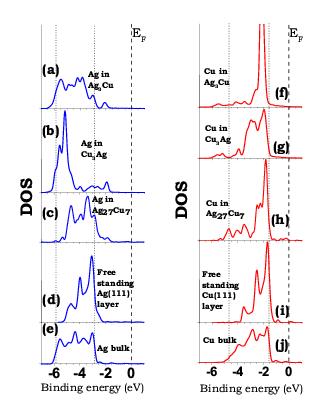}
\caption{\label{fig:ac6}
}
\end{figure*}

FIG.~\ref{fig:ac6} (Color online) Electronic DOS ($d$-band) of $Cu$ and $Ag$ atoms situated in various environments and bond length ($l$); (a) $Ag$ atoms in $Ag_{3}Cu$ ($l$ = 2.87 {\AA}); (b) $Ag$ atoms in $Cu_{3}Ag$ ($l$ = 2.70 {\AA}); (c) $Ag$ atoms in $Ag_{27}Cu_{7}$; (d) $Ag$ atoms in {\AA} free standing $Ag(111)$ monolayer ($l$ = 2.94 {\AA}); (e) $Ag$ atoms in bulk ($l$ = 2.94 {\AA}); (f) $Cu$ atoms in $Ag_{3}Cu$ ($l$ = 2.87 {\AA}); (g) $Cu$ atoms in $Cu_{3}Ag$ ($l$ = 2.70 {\AA});  (h) $Cu$ atoms in $Ag_{27}Cu_{7}$; (i) $Cu$ atoms in {\AA} free standing $Cu(111)$ monolayer ($l$ = 2.59 {\AA}); (j) $Cu$ atoms in bulk ($l$ = 2.59 {\AA}).

\begin{figure*}
\includegraphics[width=1.0\textwidth]{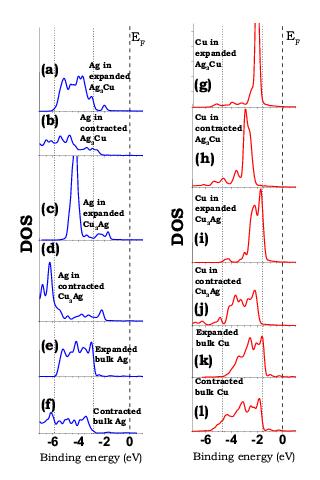}
\caption{\label{fig:ac7}
}
\end{figure*}

FIG.~\ref{fig:ac7} (Color online) Electronic DOS ($d$-band) of $Cu$ and $Ag$ atoms in pure bulk and $L1_2$ bulk $Ag-Cu$ alloys with expanded/contracted $l$: (a) $Ag$ atoms in 2.4~{\%} expanded $Ag_{3}Cu$  with $l$ = $l_{Ag_{bulk}}$ = 2.94 {\AA}; (b) $Ag$ atoms in  9.8~{\%} contracted $Ag_{3}Cu$  with $l$ = $l_{Cu_{bulk}}$ = 2.59 {\AA}; (c) $Ag$ atoms in 8.9~{\%} expanded $Cu_{3}Ag$  with $l$ = $l_{Ag_{bulk}}$; (d) $Ag$ atoms in 5.5~{\%} contracted $Cu_{3}Ag$  with $l$ = d($Cu0$-$Cu1$)= 2.55 {\AA}; (e) $Ag$ atoms 8.0~{\%} expanded bulk $Ag$; (f) $Ag$ atoms in 11.9~{\%} contracted bulk $Ag$  with $l$ = $l_{Cu_{bulk}}$; (g) $Cu$ in 2.4~{\%} expanded $Ag_{3}Cu$; (h) $Cu$ in 9.8~{\%} contracted $Ag_{3}Cu$; (i) $Cu$ in 8.9~{\%} expanded $Cu_{3}Ag$; (j) $Cu$ in 5.5~{\%} contracted $Cu_{3}Ag$; (k) $Cu$ in 8.0~{\%} expanded bulk $Cu$; (l) $Cu$ in 1.6~{\%} contracted bulk $Cu$.

\begin{figure*}
\includegraphics[width=1.0\textwidth]{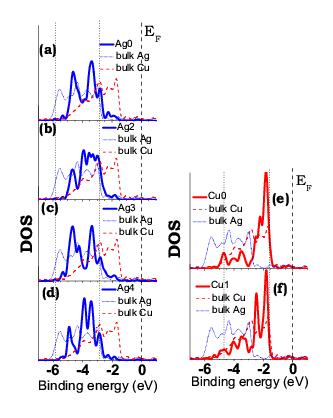}
\caption{\label{fig:ac8}
}
\end{figure*}

FIG.~\ref{fig:ac8} (Color online) The PDOS of the six non-equivalent types of atoms in the $Ag_{27}Cu_{7}$ nanoalloy is compared with the DOS of pure bulk $Cu$ (dashed lines) and $Ag$ (dotted lines). $E_F$ is shifted to 0 for all of them; (a) $Ag0$; (b) $Ag2$; (c) $Ag3$; (d) $Ag4$ (e) $Cu0$; (f) $Cu1$.

\begin{figure*}
\includegraphics[width=1.0\textwidth]{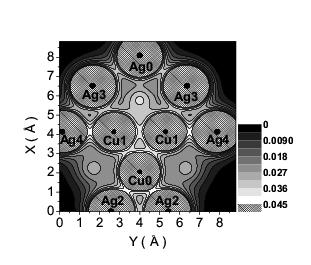}
\caption{\label{fig:ac9}
}
\end{figure*}

FIG.~\ref{fig:ac9} 2D Charge density plot at a plane that contains the 5-fold rotation axis of $Ag_{27}Cu_{7}$ nanoalloy and is, therefore, perpendicular to its mirror plane. Atoms labelled with black color are precisely centered on that plane.

\begin{figure*}
\includegraphics[width=1.0\textwidth]{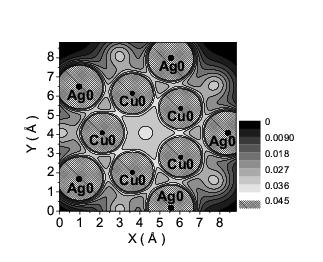}
\caption{\label{fig:ac10}
}
\end{figure*}

FIG.~\ref{fig:ac10} 2D Charge density plot at the mirror plane of $Ag_{27}Cu_{7}$. Atoms labelled with black color are precisely centered on that plane.

\begin{figure*}
\includegraphics[width=1.0\textwidth]{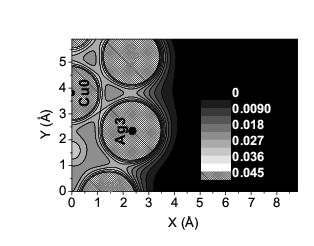}
\caption{\label{fig:ac11}
}
\end{figure*}

FIG.~\ref{fig:ac11} 2D Charge density plot at a plane that contains the positions where $Cu0$ and $Ag3$ atoms (labeled with black color) are centered. Atoms labelled with grey color are depicted in the section but are not centered precisely on that plane.

\begin{figure*}
\includegraphics[width=1.0\textwidth]{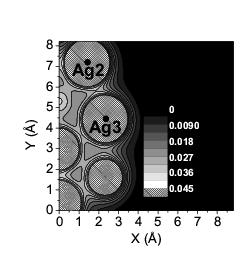}
\caption{\label{fig:ac12}
}
\end{figure*}

FIG.~\ref{fig:ac12} 2D Charge density plot at a plane that contains the positions where $Ag2$ and $Ag3$ atoms (labeled with black color) are centered. Atoms labelled with grey color are depicted in the section but are not centered precisely on that plane.

\begin{figure*}
\includegraphics[width=1.0\textwidth]{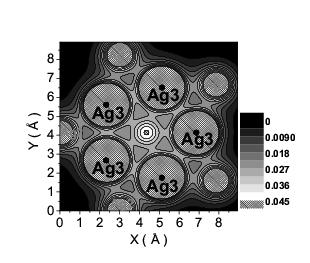}
\caption{\label{fig:ac13}
}
\end{figure*}

FIG.~\ref{fig:ac13} 2D Charge density plot at a plane which is parallel to the mirror plane of $Ag_{27}Cu_{7}$. The pentagonal layer composed of $Ag3$ atoms (labeled with black color) is contained in this plane. Atoms labelled with grey color are depicted in the section but are not centered precisely on that plane.

\begin{figure*}
\includegraphics[width=1.0\textwidth]{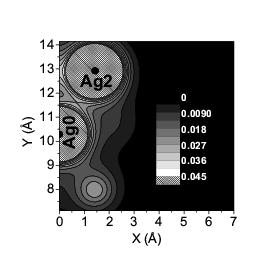}
\caption{\label{fig:ac14}
}
\end{figure*}

FIG.~\ref{fig:ac14} 2D Charge density plot at a plane that contains the positions where $Ag0$ and $Ag2$ atoms (labeled with black color) are centered. Atoms labelled with grey color are depicted in the section but are not centered precisely on that plane.

\clearpage

\clearpage

%\newpage %Just because of unusual number of tables stacked at end
%\bibliography{ALCANTARA_RAHMAN_prb}% Produces the bibliography via BibTeX.
\bibliography{ALCANTARA_RAHMAN_prb}

\begin{thebibliography}{55}
\expandafter\ifx\csname natexlab\endcsname\relax\def\natexlab#1{#1}\fi
\expandafter\ifx\csname bibnamefont\endcsname\relax
  \def\bibnamefont#1{#1}\fi
\expandafter\ifx\csname bibfnamefont\endcsname\relax
  \def\bibfnamefont#1{#1}\fi
\expandafter\ifx\csname citenamefont\endcsname\relax
  \def\citenamefont#1{#1}\fi
\expandafter\ifx\csname url\endcsname\relax
  \def\url#1{\texttt{#1}}\fi
\expandafter\ifx\csname urlprefix\endcsname\relax\def\urlprefix{URL }\fi
\providecommand{\bibinfo}[2]{#2}
\providecommand{\eprint}[2][]{\url{#2}}

\bibitem[{\citenamefont{Rossi et~al.}(2004)\citenamefont{Rossi, Rapallo,
  Mottet, Fortunelli, Baletto, and Ferrando}}]{a1}
\bibinfo{author}{\bibfnamefont{G.}~\bibnamefont{Rossi}},
  \bibinfo{author}{\bibfnamefont{A.}~\bibnamefont{Rapallo}},
  \bibinfo{author}{\bibfnamefont{C.}~\bibnamefont{Mottet}},
  \bibinfo{author}{\bibfnamefont{A.}~\bibnamefont{Fortunelli}},
  \bibinfo{author}{\bibfnamefont{F.}~\bibnamefont{Baletto}}, \bibnamefont{and}
  \bibinfo{author}{\bibfnamefont{R.}~\bibnamefont{Ferrando}},
  \bibinfo{journal}{Phys.\ Rev.\ Lett.} \textbf{\bibinfo{volume}{93}},
  \bibinfo{pages}{105503} (\bibinfo{year}{2004}).

\bibitem[{\citenamefont{Shibata et~al.}(2002)\citenamefont{Shibata, Bunker,
  Zhang, Meisel, VardemanII, and Gezelter}}]{a2}
\bibinfo{author}{\bibfnamefont{T.}~\bibnamefont{Shibata}},
  \bibinfo{author}{\bibfnamefont{B.~A.} \bibnamefont{Bunker}},
  \bibinfo{author}{\bibfnamefont{Z.}~\bibnamefont{Zhang}},
  \bibinfo{author}{\bibfnamefont{D.}~\bibnamefont{Meisel}},
  \bibinfo{author}{\bibfnamefont{C.~F.} \bibnamefont{VardemanII}},
  \bibnamefont{and} \bibinfo{author}{\bibfnamefont{J.~D.}
  \bibnamefont{Gezelter}}, \bibinfo{journal}{J.\ Am.\ Chem.\ Soc.}
  \textbf{\bibinfo{volume}{124}}, \bibinfo{pages}{11989}
  (\bibinfo{year}{2002}).

\bibitem[{\citenamefont{Darby et~al.}(2002)\citenamefont{Darby, Mortimer-Jones,
  Johnston, and Roberts}}]{a3}
\bibinfo{author}{\bibfnamefont{S.}~\bibnamefont{Darby}},
  \bibinfo{author}{\bibfnamefont{T.~V.} \bibnamefont{Mortimer-Jones}},
  \bibinfo{author}{\bibfnamefont{R.~L.} \bibnamefont{Johnston}},
  \bibnamefont{and} \bibinfo{author}{\bibfnamefont{C.}~\bibnamefont{Roberts}},
  \bibinfo{journal}{J.\ Chem.\ Phys.} \textbf{\bibinfo{volume}{116}},
  \bibinfo{pages}{1536} (\bibinfo{year}{2002}).

\bibitem[{\citenamefont{Cottancin et~al.}(2000)\citenamefont{Cottancin,
  Lerm{\'e}, Gaudry, Pellarin, Vialle, Broyer, Pr{\'e}vel, Treilleux, and
  M{\'e}linon}}]{a4}
\bibinfo{author}{\bibfnamefont{E.}~\bibnamefont{Cottancin}},
  \bibinfo{author}{\bibfnamefont{J.}~\bibnamefont{Lerm{\'e}}},
  \bibinfo{author}{\bibfnamefont{M.}~\bibnamefont{Gaudry}},
  \bibinfo{author}{\bibfnamefont{M.}~\bibnamefont{Pellarin}},
  \bibinfo{author}{\bibfnamefont{J.~L.} \bibnamefont{Vialle}},
  \bibinfo{author}{\bibfnamefont{M.}~\bibnamefont{Broyer}},
  \bibinfo{author}{\bibfnamefont{B.}~\bibnamefont{Pr{\'e}vel}},
  \bibinfo{author}{\bibfnamefont{M.}~\bibnamefont{Treilleux}},
  \bibnamefont{and}
  \bibinfo{author}{\bibfnamefont{P.}~\bibnamefont{M{\'e}linon}},
  \bibinfo{journal}{Phys.\ Rev.\ B} \textbf{\bibinfo{volume}{62}},
  \bibinfo{pages}{5179} (\bibinfo{year}{2000}).

\bibitem[{\citenamefont{Portales et~al.}(2002)\citenamefont{Portales, Saviot,
  Duval, Gaudry, Cottancin, Pellarin, Lerm{\'e}, and Broyer}}]{a5}
\bibinfo{author}{\bibfnamefont{H.}~\bibnamefont{Portales}},
  \bibinfo{author}{\bibfnamefont{L.}~\bibnamefont{Saviot}},
  \bibinfo{author}{\bibfnamefont{E.}~\bibnamefont{Duval}},
  \bibinfo{author}{\bibfnamefont{M.}~\bibnamefont{Gaudry}},
  \bibinfo{author}{\bibfnamefont{E.}~\bibnamefont{Cottancin}},
  \bibinfo{author}{\bibfnamefont{M.}~\bibnamefont{Pellarin}},
  \bibinfo{author}{\bibfnamefont{J.}~\bibnamefont{Lerm{\'e}}},
  \bibnamefont{and} \bibinfo{author}{\bibfnamefont{M.}~\bibnamefont{Broyer}},
  \bibinfo{journal}{Phys.\ Rev.\ B} \textbf{\bibinfo{volume}{65}},
  \bibinfo{pages}{165422} (\bibinfo{year}{2002}).

\bibitem[{\citenamefont{Giorgio and Henry}(2002)}]{a6}
\bibinfo{author}{\bibfnamefont{S.}~\bibnamefont{Giorgio}} \bibnamefont{and}
  \bibinfo{author}{\bibfnamefont{C.~R.} \bibnamefont{Henry}},
  \bibinfo{journal}{Eur.\ Phys.\ J.\ Appl.\ Phys.}
  \textbf{\bibinfo{volume}{20}}, \bibinfo{pages}{23} (\bibinfo{year}{2002}).

\bibitem[{\citenamefont{Moskovits et~al.}(2002)\citenamefont{Moskovits,
  Srnov{\'a}-{\v S}loufo{v\'a}, and Vl{\v c}kov{\'a}}}]{a7}
\bibinfo{author}{\bibfnamefont{M.}~\bibnamefont{Moskovits}},
  \bibinfo{author}{\bibfnamefont{I.}~\bibnamefont{Srnov{\'a}-{\v
  S}loufo{v\'a}}}, \bibnamefont{and}
  \bibinfo{author}{\bibfnamefont{B.}~\bibnamefont{Vl{\v c}kov{\'a}}},
  \bibinfo{journal}{J.\ Chem.\ Phys.} \textbf{\bibinfo{volume}{116}},
  \bibinfo{pages}{10435} (\bibinfo{year}{2002}).

\bibitem[{\citenamefont{Tada et~al.}(2002)\citenamefont{Tada, Suzuki, Ito,
  Akita, Tanaka, Kawahara, and Kobayashi}}]{a8}
\bibinfo{author}{\bibfnamefont{H.}~\bibnamefont{Tada}},
  \bibinfo{author}{\bibfnamefont{F.}~\bibnamefont{Suzuki}},
  \bibinfo{author}{\bibfnamefont{S.}~\bibnamefont{Ito}},
  \bibinfo{author}{\bibfnamefont{T.}~\bibnamefont{Akita}},
  \bibinfo{author}{\bibfnamefont{K.}~\bibnamefont{Tanaka}},
  \bibinfo{author}{\bibfnamefont{T.}~\bibnamefont{Kawahara}}, \bibnamefont{and}
  \bibinfo{author}{\bibfnamefont{H.}~\bibnamefont{Kobayashi}},
  \bibinfo{journal}{J.\ Phys.\ Chem.\ B} \textbf{\bibinfo{volume}{106}},
  \bibinfo{pages}{8714} (\bibinfo{year}{2002}).

\bibitem[{\citenamefont{Valden et~al.}(1998)\citenamefont{Valden, Lai, and
  Goodman}}]{a9}
\bibinfo{author}{\bibfnamefont{M.}~\bibnamefont{Valden}},
  \bibinfo{author}{\bibfnamefont{X.}~\bibnamefont{Lai}}, \bibnamefont{and}
  \bibinfo{author}{\bibfnamefont{D.~W.} \bibnamefont{Goodman}},
  \bibinfo{journal}{Science} \textbf{\bibinfo{volume}{281}},
  \bibinfo{pages}{1647} (\bibinfo{year}{1998}).

\bibitem[{\citenamefont{Molina and Hammer}(2003)}]{a10}
\bibinfo{author}{\bibfnamefont{L.~M.} \bibnamefont{Molina}} \bibnamefont{and}
  \bibinfo{author}{\bibfnamefont{B.}~\bibnamefont{Hammer}},
  \bibinfo{journal}{Phys.\ Rev.\ Lett.} \textbf{\bibinfo{volume}{90}},
  \bibinfo{pages}{206102} (\bibinfo{year}{2003}).

\bibitem[{\citenamefont{Haruta}(1997)}]{a11}
\bibinfo{author}{\bibfnamefont{M.}~\bibnamefont{Haruta}},
  \bibinfo{journal}{Catal.~Today} \textbf{\bibinfo{volume}{36}},
  \bibinfo{pages}{153} (\bibinfo{year}{1997}).

\bibitem[{\citenamefont{Sanchez et~al.}(1999)\citenamefont{Sanchez, Abbet,
  Heiz, Schneider, Ha1kkinen, Barnett, and Landman}}]{a12}
\bibinfo{author}{\bibfnamefont{A.}~\bibnamefont{Sanchez}},
  \bibinfo{author}{\bibfnamefont{S.}~\bibnamefont{Abbet}},
  \bibinfo{author}{\bibfnamefont{U.}~\bibnamefont{Heiz}},
  \bibinfo{author}{\bibfnamefont{W.~D.} \bibnamefont{Schneider}},
  \bibinfo{author}{\bibfnamefont{H.}~\bibnamefont{Ha1kkinen}},
  \bibinfo{author}{\bibfnamefont{R.~N.} \bibnamefont{Barnett}},
  \bibnamefont{and} \bibinfo{author}{\bibfnamefont{U.}~\bibnamefont{Landman}},
  \bibinfo{journal}{J.\ Phys.\ Chem.} \textbf{\bibinfo{volume}{103}},
  \bibinfo{pages}{9573} (\bibinfo{year}{1999}).

\bibitem[{\citenamefont{Huber et~al.}(2006)\citenamefont{Huber, Koskinen,
  H{\"a}kkinen, and Moseler}}]{a13}
\bibinfo{author}{\bibfnamefont{B.}~\bibnamefont{Huber}},
  \bibinfo{author}{\bibfnamefont{P.}~\bibnamefont{Koskinen}},
  \bibinfo{author}{\bibfnamefont{H.}~\bibnamefont{H{\"a}kkinen}},
  \bibnamefont{and} \bibinfo{author}{\bibfnamefont{M.}~\bibnamefont{Moseler}},
  \bibinfo{journal}{Nature\ Mat.} \textbf{\bibinfo{volume}{5}},
  \bibinfo{pages}{44} (\bibinfo{year}{2006}).

\bibitem[{\citenamefont{Lee et~al.}(2005)\citenamefont{Lee, Fan, Wu, and
  Anderson}}]{a14}
\bibinfo{author}{\bibfnamefont{S.}~\bibnamefont{Lee}},
  \bibinfo{author}{\bibfnamefont{C.}~\bibnamefont{Fan}},
  \bibinfo{author}{\bibfnamefont{T.}~\bibnamefont{Wu}}, \bibnamefont{and}
  \bibinfo{author}{\bibfnamefont{S.~L.} \bibnamefont{Anderson}},
  \bibinfo{journal}{J.\ Chem.\ Phys.} \textbf{\bibinfo{volume}{123}},
  \bibinfo{pages}{124710} (\bibinfo{year}{2005}).

\bibitem[{\citenamefont{Jensen}(1999)}]{a15}
\bibinfo{author}{\bibfnamefont{P.}~\bibnamefont{Jensen}},
  \bibinfo{journal}{Rev.\ Mod.\ Phys.} \textbf{\bibinfo{volume}{71}},
  \bibinfo{pages}{1695} (\bibinfo{year}{1999}).

\bibitem[{\citenamefont{Ashman et~al.}(1997)\citenamefont{Ashman, Khanna, Liu,
  Jena, Kaplan, and Mostoller}}]{a16}
\bibinfo{author}{\bibfnamefont{C.}~\bibnamefont{Ashman}},
  \bibinfo{author}{\bibfnamefont{S.~N.} \bibnamefont{Khanna}},
  \bibinfo{author}{\bibfnamefont{F.}~\bibnamefont{Liu}},
  \bibinfo{author}{\bibfnamefont{P.}~\bibnamefont{Jena}},
  \bibinfo{author}{\bibfnamefont{T.}~\bibnamefont{Kaplan}}, \bibnamefont{and}
  \bibinfo{author}{\bibfnamefont{M.}~\bibnamefont{Mostoller}},
  \bibinfo{journal}{Phys.\ Rev.\ B} \textbf{\bibinfo{volume}{55}},
  \bibinfo{pages}{15868} (\bibinfo{year}{1997}).

\bibitem[{\citenamefont{Kokko et~al.}(1990)\citenamefont{Kokko, Ojala, and
  Mansikka}}]{a17}
\bibinfo{author}{\bibfnamefont{K.}~\bibnamefont{Kokko}},
  \bibinfo{author}{\bibfnamefont{E.}~\bibnamefont{Ojala}}, \bibnamefont{and}
  \bibinfo{author}{\bibfnamefont{K.}~\bibnamefont{Mansikka}},
  \bibinfo{journal}{J.\ Phys.\:\ Condens.\ Matter}
  \textbf{\bibinfo{volume}{2}}, \bibinfo{pages}{4587} (\bibinfo{year}{1990}).

\bibitem[{\citenamefont{Doye and Calvo}(2001)}]{a18}
\bibinfo{author}{\bibfnamefont{J.~P.~K.} \bibnamefont{Doye}} \bibnamefont{and}
  \bibinfo{author}{\bibfnamefont{F.}~\bibnamefont{Calvo}},
  \bibinfo{journal}{Phys.\ Rev.\ Lett.} \textbf{\bibinfo{volume}{86}},
  \bibinfo{pages}{3570} (\bibinfo{year}{2001}).

\bibitem[{\citenamefont{von Issendorff and Cheshnovsky}(2005)}]{Issendorf}
\bibinfo{author}{\bibfnamefont{B.}~\bibnamefont{von Issendorff}}
  \bibnamefont{and}
  \bibinfo{author}{\bibfnamefont{O.}~\bibnamefont{Cheshnovsky}},
  \bibinfo{journal}{Annu. Rev. Phys. Chem.} \textbf{\bibinfo{volume}{56}},
  \bibinfo{pages}{549} (\bibinfo{year}{2005}).

\bibitem[{\citenamefont{Terakura et~al.}(1987)\citenamefont{Terakura, Oguchi,
  Mohri, and Watanabe}}]{a34}
\bibinfo{author}{\bibfnamefont{K.}~\bibnamefont{Terakura}},
  \bibinfo{author}{\bibfnamefont{T.}~\bibnamefont{Oguchi}},
  \bibinfo{author}{\bibfnamefont{T.}~\bibnamefont{Mohri}}, \bibnamefont{and}
  \bibinfo{author}{\bibfnamefont{K.}~\bibnamefont{Watanabe}},
  \bibinfo{journal}{Phys.\ Rev.\ B} \textbf{\bibinfo{volume}{35}},
  \bibinfo{pages}{2169} (\bibinfo{year}{1987}).

\bibitem[{\citenamefont{S{\'a}nchez et~al.}(1991)\citenamefont{S{\'a}nchez,
  Stark, and Moruzzi}}]{a35}
\bibinfo{author}{\bibfnamefont{J.~M.} \bibnamefont{S{\'a}nchez}},
  \bibinfo{author}{\bibfnamefont{J.~P.} \bibnamefont{Stark}}, \bibnamefont{and}
  \bibinfo{author}{\bibfnamefont{V.~L.} \bibnamefont{Moruzzi}},
  \bibinfo{journal}{Phys.\ Rev.\ B} \textbf{\bibinfo{volume}{44}},
  \bibinfo{pages}{5411} (\bibinfo{year}{1991}).

\bibitem[{\citenamefont{van-de Walle and Ceder}(2002)}]{a36}
\bibinfo{author}{\bibfnamefont{A.}~\bibnamefont{van-de Walle}}
  \bibnamefont{and} \bibinfo{author}{\bibfnamefont{G.}~\bibnamefont{Ceder}},
  \bibinfo{journal}{Rev.\ Mod.\ Phys} \textbf{\bibinfo{volume}{74}},
  \bibinfo{pages}{11} (\bibinfo{year}{2002}).

\bibitem[{\citenamefont{Heid and Bohnen}(2003)}]{r1}
\bibinfo{author}{\bibfnamefont{R.}~\bibnamefont{Heid}} \bibnamefont{and}
  \bibinfo{author}{\bibfnamefont{K.~P.} \bibnamefont{Bohnen}},
  \bibinfo{journal}{Phys.\ Rep.} \textbf{\bibinfo{volume}{387}},
  \bibinfo{pages}{151} (\bibinfo{year}{2003}).

\bibitem[{nav()}]{naval}
\eprint{Illustrations can be found in the web page
  http://cst-www.nrl.navy.mil/lattice/, provided by the Center for
  Computational Materials Science of the United States Naval Research
  Laboratory.}

\bibitem[{\citenamefont{Jones and Gunnarson}(1989)}]{r50}
\bibinfo{author}{\bibfnamefont{R.~O.} \bibnamefont{Jones}} \bibnamefont{and}
  \bibinfo{author}{\bibfnamefont{O.}~\bibnamefont{Gunnarson}},
  \bibinfo{journal}{Rev.\ Mod.\ Phys.} \textbf{\bibinfo{volume}{61}},
  \bibinfo{pages}{689} (\bibinfo{year}{1989}).

\bibitem[{\citenamefont{Baroni et~al.}()\citenamefont{Baroni, Corso,
  de~Gironcoli, Giannozzi et~al.}}]{r57}
\bibinfo{author}{\bibfnamefont{S.}~\bibnamefont{Baroni}},
  \bibinfo{author}{\bibfnamefont{A.~D.} \bibnamefont{Corso}},
  \bibinfo{author}{\bibfnamefont{S.}~\bibnamefont{de~Gironcoli}},
  \bibinfo{author}{\bibfnamefont{P.}~\bibnamefont{Giannozzi}},
  \bibnamefont{et~al.}, \eprint{Complete Quantum-ESPRESSO distribution version
  3.0 (2005) http://www.pwscf.org}.

\bibitem[{\citenamefont{Vanderbilt}(1990)}]{r53}
\bibinfo{author}{\bibfnamefont{D.}~\bibnamefont{Vanderbilt}},
  \bibinfo{journal}{Phys.\ Rev.\ B} \textbf{\bibinfo{volume}{41}},
  \bibinfo{pages}{7892} (\bibinfo{year}{1990}).

\bibitem[{\citenamefont{Perdew et~al.}(1996)\citenamefont{Perdew, Burke, and
  Ernzerhof}}]{r56}
\bibinfo{author}{\bibfnamefont{J.~P.} \bibnamefont{Perdew}},
  \bibinfo{author}{\bibfnamefont{K.}~\bibnamefont{Burke}}, \bibnamefont{and}
  \bibinfo{author}{\bibfnamefont{M.}~\bibnamefont{Ernzerhof}},
  \bibinfo{journal}{Phys.\ Rev.\ Lett.} \textbf{\bibinfo{volume}{77}},
  \bibinfo{pages}{3865} (\bibinfo{year}{1996}).

\bibitem[{\citenamefont{Methfessel and Paxton}(1989)}]{a25}
\bibinfo{author}{\bibfnamefont{M.}~\bibnamefont{Methfessel}} \bibnamefont{and}
  \bibinfo{author}{\bibfnamefont{A.~T.} \bibnamefont{Paxton}},
  \bibinfo{journal}{Phys.\ Rev.\ B} \textbf{\bibinfo{volume}{40}},
  \bibinfo{pages}{3616} (\bibinfo{year}{1989}).

\bibitem[{\citenamefont{Monkhorst and Pack}(1976)}]{a26}
\bibinfo{author}{\bibfnamefont{H.~J.} \bibnamefont{Monkhorst}}
  \bibnamefont{and} \bibinfo{author}{\bibfnamefont{J.~P.} \bibnamefont{Pack}},
  \bibinfo{journal}{Phys.\ Rev.\ B} \textbf{\bibinfo{volume}{13}},
  \bibinfo{pages}{5188} (\bibinfo{year}{1976}).

\bibitem[{\citenamefont{Baroni et~al.}(1987)\citenamefont{Baroni, Giannozzi,
  and Testa}}]{r62}
\bibinfo{author}{\bibfnamefont{S.}~\bibnamefont{Baroni}},
  \bibinfo{author}{\bibfnamefont{P.}~\bibnamefont{Giannozzi}},
  \bibnamefont{and} \bibinfo{author}{\bibfnamefont{A.}~\bibnamefont{Testa}},
  \bibinfo{journal}{Phys.\ Rev.\ Lett.} \textbf{\bibinfo{volume}{58}},
  \bibinfo{pages}{1861} (\bibinfo{year}{1987}).

\bibitem[{\citenamefont{Giannozzi et~al.}(1991)\citenamefont{Giannozzi,
  de~Gironcoli, Pavone, and Baroni}}]{r63}
\bibinfo{author}{\bibfnamefont{P.}~\bibnamefont{Giannozzi}},
  \bibinfo{author}{\bibfnamefont{S.}~\bibnamefont{de~Gironcoli}},
  \bibinfo{author}{\bibfnamefont{P.}~\bibnamefont{Pavone}}, \bibnamefont{and}
  \bibinfo{author}{\bibfnamefont{S.}~\bibnamefont{Baroni}},
  \bibinfo{journal}{Phys.\ Rev.\ B} \textbf{\bibinfo{volume}{43}},
  \bibinfo{pages}{7231} (\bibinfo{year}{1991}).

\bibitem[{\citenamefont{Blat et~al.}(1991)\citenamefont{Blat, Zein, and
  Zinenko}}]{r65}
\bibinfo{author}{\bibfnamefont{D.~K.} \bibnamefont{Blat}},
  \bibinfo{author}{\bibfnamefont{N.~E.} \bibnamefont{Zein}}, \bibnamefont{and}
  \bibinfo{author}{\bibfnamefont{V.~I.} \bibnamefont{Zinenko}},
  \bibinfo{journal}{J.\ Phys.\: Condens. Matter} \textbf{\bibinfo{volume}{3}},
  \bibinfo{pages}{5515} (\bibinfo{year}{1991}).

\bibitem[{\citenamefont{Hellwege and Hellwege}(1969)}]{r70}
\bibinfo{author}{\bibfnamefont{K.~H.} \bibnamefont{Hellwege}} \bibnamefont{and}
  \bibinfo{author}{\bibfnamefont{A.~M.} \bibnamefont{Hellwege}},
  \emph{\bibinfo{title}{Numerical data and functional relationships in science
  and technology}}, vol.~\bibinfo{volume}{2} (\bibinfo{publisher}{K. H.
  Hellwege and A. M. Hellwege; Landolt-B{\"o}rnstein-Group III Condensed
  Matter; New Series; Group III; Springer-Verlag; Berlin},
  \bibinfo{year}{1969}).

\bibitem[{com()}]{comment}
\eprint{In order to take advantage of the plane wave approach, the nanoalloy is
  modeled in a periodic supercell, which is large enough so as to minimize the
  interactions between neighboring $Ag_{27}Cu_7$ clusters by reducing the the
  overlapping between the wave functions of neighboring clusters. Valence
  electrons are thus confined spatially in the cluster, implying that electrons
  in each level specified by an index band $n$ and wave vector $\boldsymbol{k}$
  have a vanishing mean velocity given by~\cite{ash} $
  \boldsymbol{v}_n(\boldsymbol{k}) = \frac{1}{\hbar}
  \boldsymbol{\nabla}_{\boldsymbol{k}} \epsilon_n (\boldsymbol{k}) \rightarrow
  0$, implying that the electronic bands, $ \epsilon_n (\boldsymbol{k})$, are
  constant over the BZ. It is thus safe to assume that one k-point (any) is
  sufficient to accurately perform the calculation and that no additional
  information will be obtained from sampling of the BZ. To confirm the above,
  we performed total energy calculations of $Ag_{27}Cu_7$ with 1, 8, and 24
  special k-points to find that the total energy changes by only
  4$\times$10$^{-5}$ eV while forces on each atom remain below
  8$\times$10$^{-3}$ eV/{\AA}. Furthermore, non-self-consistent calculations at
  24 k-points were performed and showed that the band structure of
  $Ag_{27}Cu_7$ is a set of flat bands (see Fig.~\ref{fig:ac15})}.

\bibitem[{\citenamefont{DalCorso}(2001)}]{r80}
\bibinfo{author}{\bibfnamefont{A.}~\bibnamefont{DalCorso}},
  \bibinfo{journal}{Phys.\ Rev.\ B} \textbf{\bibinfo{volume}{64}},
  \bibinfo{pages}{235118} (\bibinfo{year}{2001}).

\bibitem[{\citenamefont{Cipriani et~al.}(2002)\citenamefont{Cipriani, Loffreda,
  Dal-Corso, S.de-Gironcoli, and Baroni}}]{a32}
\bibinfo{author}{\bibfnamefont{G.}~\bibnamefont{Cipriani}},
  \bibinfo{author}{\bibfnamefont{D.}~\bibnamefont{Loffreda}},
  \bibinfo{author}{\bibfnamefont{A.}~\bibnamefont{Dal-Corso}},
  \bibinfo{author}{\bibnamefont{S.de-Gironcoli}}, \bibnamefont{and}
  \bibinfo{author}{\bibfnamefont{S.}~\bibnamefont{Baroni}},
  \bibinfo{journal}{Surf.\ Sci.} \textbf{\bibinfo{volume}{501}},
  \bibinfo{pages}{182} (\bibinfo{year}{2002}).

\bibitem[{\citenamefont{Ferrando}()}]{a19}
\bibinfo{author}{\bibfnamefont{R.}~\bibnamefont{Ferrando}}, \eprint{Private
  Communication}.

\bibitem[{\citenamefont{Kong et~al.}(2005)\citenamefont{Kong, Li, Kong, and
  Liu}}]{a38}
\bibinfo{author}{\bibfnamefont{Y.}~\bibnamefont{Kong}},
  \bibinfo{author}{\bibfnamefont{J.~H.} \bibnamefont{Li}},
  \bibinfo{author}{\bibfnamefont{L.~T.} \bibnamefont{Kong}}, \bibnamefont{and}
  \bibinfo{author}{\bibfnamefont{B.~X.} \bibnamefont{Liu}},
  \bibinfo{journal}{Phys.\ Rev.\ B} \textbf{\bibinfo{volume}{72}},
  \bibinfo{pages}{024209} (\bibinfo{year}{2005}).

\bibitem[{\citenamefont{He et~al.}(2002)\citenamefont{He, Sheng, Lin,
  Schilling, Tittsworth, and Ma}}]{a39}
\bibinfo{author}{\bibfnamefont{J.~H.} \bibnamefont{He}},
  \bibinfo{author}{\bibfnamefont{H.~W.} \bibnamefont{Sheng}},
  \bibinfo{author}{\bibfnamefont{J.~S.} \bibnamefont{Lin}},
  \bibinfo{author}{\bibfnamefont{P.~J.} \bibnamefont{Schilling}},
  \bibinfo{author}{\bibfnamefont{R.~C.} \bibnamefont{Tittsworth}},
  \bibnamefont{and} \bibinfo{author}{\bibfnamefont{E.}~\bibnamefont{Ma}},
  \bibinfo{journal}{Phys.\ Rev.\ Lett.} \textbf{\bibinfo{volume}{89}},
  \bibinfo{pages}{125507} (\bibinfo{year}{2002}).

\bibitem[{\citenamefont{He et~al.}(2001)\citenamefont{He, Sheng, Schilling,
  Chien, and Ma}}]{a40}
\bibinfo{author}{\bibfnamefont{J.~H.} \bibnamefont{He}},
  \bibinfo{author}{\bibfnamefont{H.~W.} \bibnamefont{Sheng}},
  \bibinfo{author}{\bibfnamefont{P.~J.} \bibnamefont{Schilling}},
  \bibinfo{author}{\bibfnamefont{C.~L.} \bibnamefont{Chien}}, \bibnamefont{and}
  \bibinfo{author}{\bibfnamefont{E.}~\bibnamefont{Ma}},
  \bibinfo{journal}{Phys.\ Rev.\ Lett.} \textbf{\bibinfo{volume}{86}},
  \bibinfo{pages}{2826} (\bibinfo{year}{2001}).

\bibitem[{\citenamefont{Liu et~al.}(2001)\citenamefont{Liu, Lai, and
  Zhang}}]{a41}
\bibinfo{author}{\bibfnamefont{B.~X.} \bibnamefont{Liu}},
  \bibinfo{author}{\bibfnamefont{W.~S.} \bibnamefont{Lai}}, \bibnamefont{and}
  \bibinfo{author}{\bibfnamefont{Z.~J.} \bibnamefont{Zhang}},
  \bibinfo{journal}{Adv.\ Phys.} \textbf{\bibinfo{volume}{50}},
  \bibinfo{pages}{367} (\bibinfo{year}{2001}).

\bibitem[{\citenamefont{Peiner and Kopitzki}(1988)}]{a42}
\bibinfo{author}{\bibfnamefont{E.}~\bibnamefont{Peiner}} \bibnamefont{and}
  \bibinfo{author}{\bibfnamefont{K.}~\bibnamefont{Kopitzki}},
  \bibinfo{journal}{Nucl.\ Instrum.\ Methods\ Phys.\ Res.}
  \textbf{\bibinfo{volume}{34}}, \bibinfo{pages}{173} (\bibinfo{year}{1988}).

\bibitem[{\citenamefont{Kara and Rahman}(2005)}]{a43}
\bibinfo{author}{\bibfnamefont{A.}~\bibnamefont{Kara}} \bibnamefont{and}
  \bibinfo{author}{\bibfnamefont{T.~S.} \bibnamefont{Rahman}},
  \bibinfo{journal}{Surf.\ Sci.\ Rep.} \textbf{\bibinfo{volume}{56}},
  \bibinfo{pages}{159} (\bibinfo{year}{2005}).

\bibitem[{\citenamefont{Bogdanoff et~al.}(1999)\citenamefont{Bogdanoff, Fultz,
  and Rosenkranz}}]{a45}
\bibinfo{author}{\bibfnamefont{P.~D.} \bibnamefont{Bogdanoff}},
  \bibinfo{author}{\bibfnamefont{B.}~\bibnamefont{Fultz}}, \bibnamefont{and}
  \bibinfo{author}{\bibfnamefont{S.}~\bibnamefont{Rosenkranz}},
  \bibinfo{journal}{Phys.\ Rev.\ B} \textbf{\bibinfo{volume}{60}},
  \bibinfo{pages}{3976} (\bibinfo{year}{1999}).

\bibitem[{\citenamefont{Sharma and Singh}(1971)}]{a44}
\bibinfo{author}{\bibfnamefont{P.~K.} \bibnamefont{Sharma}} \bibnamefont{and}
  \bibinfo{author}{\bibfnamefont{N.}~\bibnamefont{Singh}},
  \bibinfo{journal}{Phys.\ Rev.\ B} \textbf{\bibinfo{volume}{4}},
  \bibinfo{pages}{4636} (\bibinfo{year}{1971}).

\bibitem[{\citenamefont{Schmidt et~al.}(1998)\citenamefont{Schmidt, Kusche, von
  Issendorff, and Haberland}}]{a46}
\bibinfo{author}{\bibfnamefont{M.}~\bibnamefont{Schmidt}},
  \bibinfo{author}{\bibfnamefont{R.}~\bibnamefont{Kusche}},
  \bibinfo{author}{\bibfnamefont{B.}~\bibnamefont{von Issendorff}},
  \bibnamefont{and}
  \bibinfo{author}{\bibfnamefont{H.}~\bibnamefont{Haberland}},
  \bibinfo{journal}{Nature} \textbf{\bibinfo{volume}{393}},
  \bibinfo{pages}{238} (\bibinfo{year}{1998}).

\bibitem[{\citenamefont{Hild et~al.}(1998)\citenamefont{Hild, Dietrich,
  Kr{\"u}ckeberg, Lindinger, L{\"u}tzenkirchen, Schweikhard, Walther, and
  Ziegler}}]{a48}
\bibinfo{author}{\bibfnamefont{U.}~\bibnamefont{Hild}},
  \bibinfo{author}{\bibfnamefont{G.}~\bibnamefont{Dietrich}},
  \bibinfo{author}{\bibfnamefont{S.}~\bibnamefont{Kr{\"u}ckeberg}},
  \bibinfo{author}{\bibfnamefont{M.}~\bibnamefont{Lindinger}},
  \bibinfo{author}{\bibfnamefont{K.}~\bibnamefont{L{\"u}tzenkirchen}},
  \bibinfo{author}{\bibfnamefont{L.}~\bibnamefont{Schweikhard}},
  \bibinfo{author}{\bibfnamefont{C.}~\bibnamefont{Walther}}, \bibnamefont{and}
  \bibinfo{author}{\bibfnamefont{J.}~\bibnamefont{Ziegler}},
  \bibinfo{journal}{Phys.\ Rev.\ A} \textbf{\bibinfo{volume}{57}},
  \bibinfo{pages}{2786} (\bibinfo{year}{1998}).

\bibitem[{TSR({\natexlab{a}})}]{TSR1}
\eprint{H. Yildirim, K. Kara, and T. S. Rahman, to be published}.

\bibitem[{TSR({\natexlab{b}})}]{TSR2}
\eprint{Thermodynamics of finite quantum-mechanical interacting systems depends
  on the number $\Gamma$ $-$ as a function of the energy, $E$, number of
  particles, $N$, and volume, $V$ $-$ of distinct eigenstates of the
  Hamiltonian of the system that render a total energy $E'= E \pm \Delta $,
  where $\Delta << E$. The configurational entropy is then given by $S_{conf} =
  k_B ln(\Gamma)$. While obtaining the exact solution of the interacting
  Hamiltonian is a formidable task, we can obtain $S_{conf}(\Gamma)$ directly
  for an interacting bimetallic $N$-atom cluster $X_{N1}Y_{N-N1}$. Namely, the
  total energy and the electronic structure of a $X_{N1}Y_{N-N1}$ cluster are
  defined by the geometric configuration of its nuclei and, hence, different
  geometric configurations of the atoms define distinct eigenstates that may or
  may not be degenerate. We can thus say straightforwardly that a cluster
  $X_{N1}Y_{N-N1}$ with $m$ degenerate isomers will have a configurational
  entropy, $S_{conf} = k_Bln(m)$. Rossi et al., using tight binding many-body
  potentials in combination with the GGO method, found that the total energy of
  the next-lowest-energy isomer has a total energy 0.3 eV higher than that of
  $Ag_{27}Cu_7$. DFT, in turn, finds that such an isomer has an energy 0.8 eV
  higher than that of $Ag_{27}Cu_7$. In this case, therefore, we may assume
  that $Ag_{27}Cu_7$ has no degenerate isomers and so $S_{conf}$ is zero. For
  the other clusters in this family, however, a similar investigation of the
  isomers has to be performed, keeping in mind that even for doubly degenerate
  isomers ($m$=2), configurational entropic contributions to the free energy of
  $TS_{conf}$ = 18 meV (T=300 K) will arise and give rise to clusters with free
  energy lower than $Ag_{27}Cu_7$, at room temperature.}

\bibitem[{\citenamefont{Kokko}(1999)}]{a49}
\bibinfo{author}{\bibfnamefont{K.}~\bibnamefont{Kokko}}, \bibinfo{journal}{J.\
  Phys.\ Condens.\ Matter} \textbf{\bibinfo{volume}{11}}, \bibinfo{pages}{6685}
  (\bibinfo{year}{1999}).

\bibitem[{\citenamefont{S{\'a}nchez et~al.}(2003)\citenamefont{S{\'a}nchez,
  Leiva, and Schmickler}}]{a52}
\bibinfo{author}{\bibfnamefont{C.~G.} \bibnamefont{S{\'a}nchez}},
  \bibinfo{author}{\bibfnamefont{E.~P.~M.} \bibnamefont{Leiva}},
  \bibnamefont{and}
  \bibinfo{author}{\bibfnamefont{W.}~\bibnamefont{Schmickler}},
  \bibinfo{journal}{Electrochem. Comm.} \textbf{\bibinfo{volume}{5}},
  \bibinfo{pages}{584} (\bibinfo{year}{2003}).

\bibitem[{\citenamefont{Wertheim et~al.}(1986)\citenamefont{Wertheim, DiCenzo,
  and Buchanan}}]{a50}
\bibinfo{author}{\bibfnamefont{G.~K.} \bibnamefont{Wertheim}},
  \bibinfo{author}{\bibfnamefont{S.~B.} \bibnamefont{DiCenzo}},
  \bibnamefont{and} \bibinfo{author}{\bibfnamefont{D.~N.~E.}
  \bibnamefont{Buchanan}}, \bibinfo{journal}{Phys.\ Rev.\ B}
  \textbf{\bibinfo{volume}{33}}, \bibinfo{pages}{5384} (\bibinfo{year}{1986}).

\bibitem[{\citenamefont{Baletto et~al.}(2003)\citenamefont{Baletto, Mottet, and
  Ferrando}}]{a37}
\bibinfo{author}{\bibfnamefont{F.}~\bibnamefont{Baletto}},
  \bibinfo{author}{\bibfnamefont{C.}~\bibnamefont{Mottet}}, \bibnamefont{and}
  \bibinfo{author}{\bibfnamefont{R.}~\bibnamefont{Ferrando}},
  \bibinfo{journal}{Eur.\ Phys.\ J.\ D} \textbf{\bibinfo{volume}{24}},
  \bibinfo{pages}{233} (\bibinfo{year}{2003}).

\bibitem[{\citenamefont{Ashcroft and Mermin}(1976)}]{ash}
\bibinfo{author}{\bibfnamefont{N.~W.} \bibnamefont{Ashcroft}} \bibnamefont{and}
  \bibinfo{author}{\bibfnamefont{N.~D.} \bibnamefont{Mermin}},
  \emph{\bibinfo{title}{Solid State Physics}}, vol. \bibinfo{volume}{1st Ed.}
  (\bibinfo{publisher}{Thomson Learning, Inc.}, \bibinfo{year}{1976}).

\end{thebibliography}
\end{document}